\documentclass[twocolumn]{aastex61}

\accepted{10/10/17}

\submitjournal{ApJ}

\shorttitle{Climate of 55 Cancri e}
\shortauthors{Hammond \& Pierrehumbert}
\begin{document}

\title{Linking the Climate and Thermal Phase Curve of 55 Cancri e}

\correspondingauthor{Mark Hammond}
\email{mark.hammond@physics.ox.ac.uk}

\author{Mark Hammond}
\affiliation{University of Oxford}
\&
\author{Raymond Pierrehumbert}
\affiliation{University of Oxford}

%%%%%%%%%%%%%%%%%%%%%%%%%%%%%%%%%%%%%%%%%%%%%%%%%%%%%%%%%%%%%%%%%%%%%%%%%%

\begin{abstract}
The thermal phase curve of 55 Cancri e is the first measurement of the temperature distribution of a tidally locked Super-Earth, but raises a number of puzzling questions about the planet's climate. The phase curve has a high amplitude and peak offset, suggesting that it has a significant eastward hot-spot shift as well as a large day-night temperature contrast. We use a general circulation model to model potential climates, and investigate the relation between bulk atmospheric composition and the magnitude of these seemingly contradictory features. We confirm theoretical models of tidally locked circulation are consistent with our numerical model of 55 Cnc e, and rule out certain atmospheric compositions based on their thermodynamic properties. Our best-fitting atmosphere has a significant hot-spot shift and day-night contrast, although these are not as large as the observed phase curve. We discuss possible physical processes which could explain the observations, and show that night-side cloud formation from species such as SiO from a day-side magma ocean could potentially increase the phase curve amplitude and explain the observations. We conclude that the observations could be explained by an optically thick atmosphere with a low mean molecular weight, a surface pressure of several bar and a strong eastward circulation, with night-side cloud formation a possible explanation for the difference between our model and the observations.

\end{abstract}


%%%%%%%%%%%%%%%%%%%%%%%%%%%%%%%%%%%%%%%%%%%%%%%%%%%%%%%%%%%%%%%%%%%%%%%%%%

\section{Introduction}\label{sec:intro}
The first phase curve of a Super-Earth was measured by \citet{demory2016map} using Spitzer observations of 55 Cancri e, following the measurement of transits in the visible \citep{winn2011transit} and infrared \citep{demory2011detection}. 55 Cnc e is a Super-Earth discovered by \citet{mcarthur2004discovery} with mass 8.63 M\textsubscript{$\Earth$} and radius 2.00 R\textsubscript{$\Earth$} in a close, tidally locked orbit with period 0.737 days. The thermal phase curve has a large amplitude and an offset between its secondary eclipse and its phase maximum. \citet{demory2016map} used the curve to reconstruct a temperature map with a maximum hemisphere-averaged $4.5 \micron$ brightness temperature of $(2700 \pm 270)\ \mathrm{K}$, day-night contrast of $(1300 \pm 670)\ \mathrm{K}$, and a hot-spot shifted eastwards by $(41 \pm 12)\degr$.

55 Cnc e is a member of a class of planets known as ``lava planets'' which are in such close orbits that they are likely to be tide-locked and have a permanent day-side magma ocean. It has been argued that the atmospheres of such planets could consist of thin mineral-vapour atmospheres outgassed from the magma ocean \citep{leger2011corot} \citep{castan2011hot}. Such thin atmospheres, consisting of a few millibar or less surface pressure, cannot transport much heat apart from possible lateral heat redistribution within the magma ocean, so would yield a phase curve very similar to that of an airless rocky planet with a very cold night-side such as discussed by \citet{maurin2012thermal}.

The thermal phase curve of 55 Cnc e presents the possibility of testing this picture of lava planets, and in particular to determine whether the phase curve demands the presence of a thick noncondensible background atmosphere. In this paper, we use a general circulation model (GCM) to model a range of hypothetical climates for 55 Cnc e and reconstruct their thermal phase curve, in order to test whether the observed phase curve is inconsistent with the presence of a thick atmosphere. We explore which atmospheric compositions are compatible with the phase curve. The results we have obtained for 55 Cnc e will carry over readily to the interpretation of other lava planet phase curves when they become available. The general utility of thermal phase curves in determining characteristics of exoplanets and their atmospheres has been discussed in \citet{selsis2011thermal} and \citet{maurin2012thermal}.

The transit depth spectra reported in \citet{tsiaras2016detection} require a thick $\mathrm{H_2}$-rich atmosphere. However, \citet{lammer2013blowoff} calculated that an H\textsubscript{2} atmosphere on 55 Cnc e would have a hydrodynamic escape rate of up to $2.8 \times 10^{9}\ \mathrm{gs}^{-1}$. This implies that a 10 bar atmosphere would be lost in less than one million years, making it implausible that an $\mathrm{H_{2}}$-rich atmosphere could be maintained on this planet.   However, the study of exoplanets has yielded up many objects that according to previous conceptions should not exist, so in this paper we will take the idea of an $\mathrm{H_{2}}$-rich atmosphere seriously, and ask what features of the phase curve measured by \citet{demory2016map} are compatible with, or demand, a low molecular weight atmosphere.

In order to focus on dynamical behavior in this initial study, we make a number of simplifying assumptions regarding the radiative behavior of the atmosphere. First, we assume the atmosphere to be transparent to incoming stellar radiation, so that all of the shortwave radiation is absorbed at the ground, leading to a deep day-side convective layer.  This assumption is based on estimates of the shortwave opacity of likely cloud-free atmospheres of up to 10 bars. The addition of a small amount of shortwave opacity would not change our results much, so long as atmospheric absorption occurs near enough the surface to drive a convective troposphere.  Very thick shortwave-opaque atmospheres could instead have a deep radiative-equilibrium layer with a thin dynamically active layer near the top; we shall not consider such atmospheres in the present paper.

In the infrared, the atmosphere is assumed to act as a grey gas with specified
optical thickness $\tau_{\infty}$ and opacity $\kappa$. This is not inconsistent with the assumption of an atmosphere largely transparent to incoming stellar radiation, because 55 Cancri is a G star, with a relatively low proportion of its output in the near-IR. The use of gray gas radiation for climate calculations is not a serious source of inaccuracy as the circulation is primarily affected by the radiation scheme via the surface temperature relative to the radiating temperature of the planet. The optical thickness $\tau_{\infty}$ can be tuned to match the temperature that would be yielded by an assumed real-gas atmosphere, so in this paper we use $\tau_{\infty}$ primarily as a way to control surface temperature.

Non-grey radiative effects are taken into account when we interpret the results in terms of the corresponding Spitzer $4.5 \micron$ phase curve, in that we consider the emission from a range of different atmospheric levels and not just the grey radiating level. This allows for the possibility that the atmospheric composition may support an infrared window region near $4.5 \micron$, allowing radiation from deeper in the atmosphere, or a source of anomalous opacity (e.g. clouds) there, forcing the radiating level to be higher in the atmosphere.

The surface pressure determines the atmospheric mass via the hydrostatic relation. For a given surface pressure and $\tau_{\infty}$, atmospheric composition affects the climate through mean molecular weight and specific heat. However, molar specific heat is only weakly dependent on composition, because it is primarily determined by the number of active degrees of freedom. For example, at 2000K the molar specific heats of CO, $\mathrm{N_2}$ and $\mathrm{H_2}$ vary by no more than 3.4\% relative to the mean value of 35.5 $\mathrm{J mol^{-1} K^{-1}}$, with similar results for other common diatomic gases. Triatomic gases have only a modestly greater molar specific heat at 2000K, e.g. 60\% higher for $\mathrm{CO_2}$ or 46\% higher for $\mathrm{H_2O}$. Therefore, the heat capacity of the atmosphere, which largely determines an atmosphere's ability to transport heat, is mostly determined by surface pressure and molecular weight. The mean molecular weight also affects the speed of gravity waves in the atmosphere, through its influence on the gas constant. This speed determines the character of many atmospheric waves which directly transport heat and are implicated in the generation of super-rotating low-latitude jets, which also transport heat. We present our simulation results in terms of a range of H\textsubscript{2}-N\textsubscript{2} mixtures, but they would apply accurately to any other diatomic mixture with the same molecular weight, and with only moderate inaccuracy to triatomic-dominated mixtures.

The measured phase curve of 55 Cnc e exhibits two features that demand substantial horizontal heat transport.  First, the hot spot of the planet is shifted 41$\degr$  eastward relative to the substellar point. Second, the night-side temperature of the planet is quite high -- on the order of 1300K -- demanding delivery of $1.6 \times 10^5 \mathrm{W/m^2}$ of heating to maintain it. However, the day-night temperature difference is also large -- on the order of 1300K -- which puts a limit on the efficiency of the heat transporting mechanism.

It has been suggested that the implied heat transport on 55 Cnc e might be carried by the magma ocean.  However, \citet{kite2016atmosphere} argued that a magma ocean could not redistribute enough heat to affect a planet's measured phase curve. It is also conceivable that tidal heating could contribute to maintaining the night-side temperature.  In this paper, we will focus on the question of whether atmospheric heat transport alone can account for the phase curve, though we will offer some remarks in Section \ref{sec:discussion} on problems with tidal heating as an explanation of the night-side temperature.

The hot-spot phase shift and phase curve amplitude on tide-locked planets have been extensively studied in connection with interpretation of Hot Jupiter phase curves. For sufficiently short period orbits, the global circulation of such atmospheres is dominated by the effects of planetary scale equatorial Rossby and Kelvin waves which drive a superrotating jet (\citet{showman2011equatorial}, \citet{heng2015atmospheric}). The circulation system transports heat eastwards around the equator, shifting the hot-spot from the substellar point and warming the night-side of the planet. The observed phase curve of 55 Cnc e poses the particular challenge that its large 41$\degr$ hot-spot shift suggests strong heat redistribution, but its large 1300 K day-night difference suggests weak heat redistribution.  The need to negotiate the tension between these two requirements puts strong constraints on the kind of atmosphere the planet can have.

The analysis in this paper builds on the results of \citet{cowan2011statistics}, \citet{menou2012scaling}, and \citet{komacek2016atmospheric}, who explored the effect of parameters such as the mean molecular weight on the thermal phase curve of Hot Jupiters via the radiative and advective timescales.  \citet{koll2015deciphering} modelled the relation between atmospheric properties and broadband thermal phase curve for terrestrial planets in a regime (expected to be appropriate to most tidally locked planets) with a significant phase curve amplitude, but very little hot-spot offset.  We have identified a regime which can support both a notable amplitude and offset.

Section \ref{sec:model} describes our model and explains the physical processes which it includes. Section \ref{sec:theory} lays out the current theory of global circulation on tidally locked planets, where we discuss the key nondimensional parameters and situate 55 Cnc e in the space of circulation regimes.  We describe the results from our experiments in Section \ref{sec:results}, focusing on their temperature distributions and simulated phase curves in comparison to the results of \citet{demory2016map}. Further interpretation of the results is provide in Section \ref{sec:discussion} and our principal findings are summarized in Section \ref{sec:conclusions}. Our best-fit clear-sky atmosphere has a surface pressure of 5 bar and a mean molecular weight of $4.6\ \mathrm{gmol}^{-1}$. This molecular weight would support the hypothesis of an $\mathrm{H_2}$-rich atmosphere; however, it is the observed hot-spot phase shift which favours low molecular weight, underscoring the importance of accurate measurements of this quantity for future observations of 55 Cnc e and other lava planets. A diagnostic estimate of cloud effects indicates that Na clouds would not form in such an atmosphere, but that SiO clouds could form on the night-side and bring the modeled night-side brightness temperature more in line with observations. Our results on the vertical structure of the temperature pattern underscore the importance of future measurements of spectrally resolved phase curves for 55 Cnc e and other lava planets, which would provide an important window into atmospheric composition and dynamics.

%%%%%%%%%%%%%%%%%%%%%%%%%%%%%%%%%%%%%%%%%%%%%%%%%%%%%%%%%%%%%%%%%%%%%%%%%%

\section{Model}\label{sec:model}
We modelled the atmosphere of 55 Cnc e using Exo-FMS, an idealised general circulation model (GCM) based on the finite-volume dynamical core of the software framework FMS described by \citet{lin2004vertically}. Other GCMs using the FMS framework have been used to model terrestrial, tidally locked exoplanets by \citet{merlis2010atmospheric}, \citet{heng2011atmospheric}, \citet{koll2015deciphering}, and \citet{koll2016temperature}. Other GCMs have been used to model tidally locked super-Earths by \citet{carone2014connecting}, \citet{kataria2014atmospheric}, and \citet{charnay20153d}.Our model planet has radius $r_{p} = 1.91\ r_{\Earth}$, orbital period $P = 0.737\ \mathrm{days}$, surface gravity $g = 21.7\ \mathrm{ms}^{-2}$ \citep{demory2016map}, and incoming stellar flux $3.55 \times 10^{6}\ \mathrm{Wm}^{-2}$ \citep{von201155}.

Our model is the same as that used in \citet{pierrehumbert2016dynamics}, but with a dry atmosphere (no condensable species). It uses the 3D fluid-dynamical core on a 144x96x40 grid, a 1D grey-gas radiative solver, and a 1D dry-convective adjustment routine. The top pressure level is $10^{-5}p_{s}$, for surface pressure p\textsubscript{s}. The model solves the primitive equations, then calculates the radiative fluxes and heating in each 1D column of the grid and updates their temperature, then adjusts any unstable parts of each column towards the dry adiabat. We use two-stream grey-gas radiative transfer so as to focus on key dynamical aspects of the problem. At temperatures as high as that of 55 Cnc e, a small portion of the emitted radiation has short enough wavelength that it can be affected by Rayleigh scattering, but we neglect this effect.

The specific calculations reported here are for atmospheres with the thermodynamic properties of H\textsubscript{2}-N\textsubscript{2} mixtures, though as noted in the introduction the composition affects the thermodynamic properties primarily through the  mean molecular weight of the mixture.  The actual composition has a much stronger effect on the radiative properties of the atmosphere, captured in our simulations by the specified infrared optical thickness $\tau_{\infty}$. To provide a point of reference, we carried out 1D radiative-convective simulations for 55 Cnc e for a 10 bar pure $\mathrm{H_2}$ atmosphere, using both a realistic representation of the $\mathrm{H_2}$ collisional opacity (based on \citep{pierrehumbert2011hydrogen}) and a grey-gas approximation. It was found that the surface temperature and vertical structure were nearly identical between the real-gas calculation and a grey-gas calculation with grey optical depth $\tau_{\infty} = 8$, so the surface temperature we quote for this case can be considered realistic for a pure 10 bar $\mathrm{H_2}$ atmosphere. This corresponds to an opacity of  22.4 cm\textsuperscript{2}kg\textsuperscript{-1}, which we used in all of our tests (apart from those in Section \ref{sec:tauinf_effect}) in order to compare the effect of other parameters. In reality, $\kappa$ should scale quadratically with pressure if it were only due to collision-induced absorption. Given the range of possible constituents that could contribute to the infrared opacity of the atmosphere, we treat $\tau_{\infty}$ as an independent parameter of the atmosphere; this helps to isolate the dynamical vs. radiative effects of composition.

In the model, the outgoing $4.5\ \micron$ radiation is radiated from the upper reaches of the atmosphere in all our tests, as they all have a high grey-gas optical thickness. In reality, the real-gas absorption spectrum might have a stronger or weaker opacity at $4.5\ \micron$ than the assumed ``average'' of the grey-gas approximation. This means that $4.5\ \micron$ radiating level relevant to the Spitzer observations of \citet{demory2016map} could be higher or lower in the atmosphere than in our results, so we will consider the temperature distribution at multiple levels of the atmosphere and not just at the grey-gas radiating level.

We neglected atmospheric absorption of incoming stellar flux, as our 1D real-gas radiative-convective model showed that the H\textsubscript{2} shortwave absorption is too weak to greatly affect the temperature profiles (and N\textsubscript{2} shortwave absorption is even weaker). Given high infrared opacity, introduction of a moderate shortwave opacity would not significantly change our results, as only a small amount of shortwave radiation needs to reach the surface in order to maintain a deep convective troposphere, as is exemplified by the case of Venus \citep[Ch. 4]{ClimateBook}.   Introduction of a very strong upper-atmosphere shortwave absorber, such as could be caused by some types of clouds,  would fundamentally alter the picture by creating a deep non-convective isothermal layer extending from the surface to the absorbing layer.  We do not consider such situations in this paper, but the possibility needs to be kept in mind.

We set the surface albedo to zero and neglected the scattering albedo of the atmosphere. This means that any temperatures from our results are an upper bound, as the absorbed stellar flux will be lower. However, a low albedo seems likely for the planet given the observed high brightness temperature.

The models were initialised with zero wind speed and the same temperature profile in each vertical column, which had a specified surface temperature and followed the dry adiabat up to a certain temperature, where it was set to an isotherm. \citet{liu2013atmospheric} showed that the results of models of tidally locked Hot Jupiter atmospheres are insensitive to initial conditions, so we do not expect our results to be sensitive to this choice of initialisation.

Following \citet{boutle2017proxima} we regarded the circulation to have reached equilibrium when the top of atmosphere radiation budget is in balance and the stratospheric temperature stops evolving.  We also checked that the global mean temperature and the winds had stopped evolving.  In contrast to the 300 days it typically took the simulations of the tidally locked planet Proxima b in \citet{boutle2017proxima} to equilibrate, our simulations had generally reached equilibrium after 10 Earth days, owing to the short radiative time scale associated with the high temperature of a lava planet. All results presented are for averages over the final 10 days of 50 day runs.

%%%%%%%%%%%%%%%%%%%%%%%%%%%%%%%%%%%%%%%%%%%%%%%%%%%%%%%%%%%%%%%%%%%%%%%%%%

\section{Scaling theory for tidally locked planetary circulation}\label{sec:theory}

\begin{figure}
\plotone{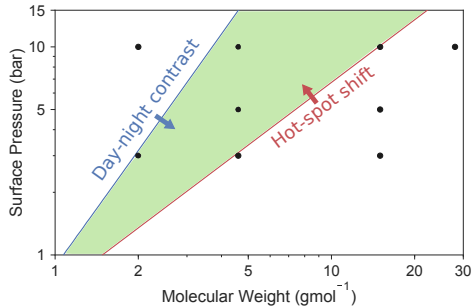}
\caption{Our parameter space, where the green region is the regime predicted to support both a significant hot-spot shift and day-night contrast. The black points show the atmospheres we tested, detailed in Table \ref{tab:paramstable}. We used a fixed advection speed 1000 ms\textsuperscript{-1} and fixed mean temperature 2000K. The lines correspond to a predicted hot-spot shift of 20$\degr$ and a day-night contrast of 80\% of the equilibrium contrast (with no heat transport). We tested some of the points with multiple values of $\tau_{\infty}$ -- in our parameter space, we do not expect it to affect the hot-spot shift or fractional day-night contrast (see Sections \ref{sec:theory} and \ref{sec:tauinf_effect}). \label{fig:param_map}}
\end{figure}

We will now discuss the characteristic time  and length scales, and corresponding nondimensional parameters, that govern a planet's heat-transport properties. The fundamental externally imposed velocity scale is the external gravity wave speed $U_{wave} = \sqrt{\frac{R^* T}{\mu}}$, where $\mu$ is the mean molecular weight of the atmosphere, $R^*$ is the universal gas constant, and $T$ is the characteristic temperature of the dynamically active layer of the atmosphere, assumed here to be on the same order as the mean radiating temperature of the planet.  \citet{zhang2016effects} use a somewhat different expression to estimate $U_{wave}$, but the result differs only by order an order unity constant from the one used here.

From $U_{wave}$ we can build the global radius of deformation
\begin{equation}
L_{d} \equiv \frac{\sqrt{R^{*}T/\mu}}{\Omega}
\end{equation}
and the wave timescale
\begin{equation}
t_{wave} \equiv \frac{r_p}{U_{wave}} = r_p \sqrt{\frac{\mu}{R^{*} T}}
\end{equation}
where  $\Omega$ is the angular velocity of the planet's spin ($2\pi/P$ for the tide locked case). A key nondimensional parameter governing the dynamical regime is then
\begin{equation}
\Lambda \equiv \frac{L_{d}}{r_{p}} = \frac{1}{\Omega t_{wave}}
\end{equation}
$\Lambda$ is the Weak Temperature Gradient (WTG) parameter discussed in \citet{pierrehumbert2016dynamics} , and constitutes a nondimensional measure of the importance of the planet's rotation.  Large $\Lambda$ corresponds to a slowly rotating planet, in which the Coriolis force is too week to support large temperature gradients. Such planets (with some caveats, discussed in \citet{pierrehumbert2016dynamics}) tend to have globally weak horizontal temperature gradients. Planets with $\Lambda$ order unity or smaller are more Earthlike, with strong meridional (north-south) temperature gradients and strong zonal (east-west) jets.  Such planets can still have small zonal temperature gradients owing to heat transport by the jets, but the magnitude and phase shift of such variations depends on the magnitude of radiative damping.

For 55 Cnc e, using a typical temperature of 2500K, $\Lambda$ ranges from 2.7 for a pure $\mathrm{H_2}$ atmosphere to 0.72 for a pure $\mathrm{N_2}$ atmosphere.  To put these values into perspective, $\Lambda = 0.6$ for Earth, which exhibits strong midlatitude temperature gradients but weak tropical temperature gradients, though it should be kept in mind that tide-locked planets have a stronger zonal variation in stellar heating than does Earth.  Hot Jupiters have a similar $L_d$ to 55 Cnc e, but much larger radius, and therefore have much smaller $\Lambda$ and are correspondingly more strongly influenced by rotation.  Very low molecular weight atmospheres on 55 Cnc e edge more into the WTG regime, with the dynamical effects of rotation becoming more dominant as molecular weight is increased, but any atmosphere  the planet might have will put it in a dynamical regime that is strongly influenced by rotation.

One can also define a time scale based on the characteristic jet speed $U$ on the planet
\begin{equation}
t_{adv} \equiv \frac{r_{p}}{U}
\end{equation}
$U$ is not an externally imposed parameter, because it is an emergent property determined by the externally imposed planetary parameters (instellation, atmospheric composition, size, etc.). For rapidly rotating tide-locked planets, the jets primarily take the form of equatorial super-rotating (eastward) circulations.  Because of the difficulty of estimating the jet speed {\it a priori} in terms of the planetary parameters, $U$ is sometimes chosen based on general circulation model simulations, in which case scalings based on $t_{adv}$ become diagnostic rather than predictive. Alternatively, insofar as phase curve properties can be shown to depend on $t_{adv}$, fits of theoretical to observed phase curves can be used to estimate the jet speed. \citet{zhang2016effects} suggest the equilibrium cyclostrophic wind as a reasonable rough estimate of jet speed; it can be shown that this estimate has the same order of magnitude as $U_{wave}$.

Most studies of the phase curve properties have focused on the radiative damping time scale, $t_{rad}$ relative to $t_{adv}$ or $t_{wave}$ as the key nondimensional parameter.
For a planet with a deep convective troposphere heated by stellar absorption near the surface,
\begin{equation}
t_{rad} \approx \frac{p_{s}}{\mu g}  \frac{c_{p,mol}}{4 \sigma T_{rad}^3}
\end{equation}
where $p_s$ is the surface pressure and $c_{p,mol}$ is the molar specific heat.  $T_{rad}$ is the mean radiating temperature of the planet, which is fixed by the planet's net absorption of stellar energy and does not depend on the atmospheric opacity or optical thickness. The pressure used in this expression differs from that in the estimate in \citet{zhang2016effects} because we consider planets with a deep convective troposphere, as opposed to fluid planets like hot Jupiters or GJ1214b where the thermal emission almost inevitably comes from a limited radiative-equilibrium layer near the top of the atmosphere. For a given planet, $t_{rad}$ varies linearly in proportion to $p_s/\mu$, so that increasing surface pressure increases the damping time in the same way as decreasing mean molecular weight. $t_{rad}/t_{wave}$ provides the fundamental {\it a priori} measure of the ability of an atmosphere to transport enough heat to even out the day-night temperature difference, though in some cases the diagnostic quantity $t_{rad}/t_{adv}$ could prove more accurate, in circumstances when $t_{adv}$ is significantly shorter than $t_{wave}$. Letting the transport time $t_{trans}$ be the lesser of $t_{adv}$ and $t_{wave}$, if they differ significantly, we can identify three regimes:
\begin{enumerate}
  \item $t_{trans} >> t_{rad}$, where air cools faster than it travels, leading to a strong day-night contrast and a weak hot-spot shift, i.e. a phase curve with a large amplitude and small offset.
  \item $t_{trans} << t_{rad}$, where air travels faster than it cools, leading to a weak day-night contrast and a large hot-spot shift, i.e. a phase curve with a small amplitude and large offset.
  \item $t_{trams} \sim t_{rad}$, where air cools on the same timescale as it circulates. This may lead to both a significant hot-spot shift and day-night contrast, i.e. a phase curve with a large amplitude and peak offset like the curve of 55 Cnc e.
\end{enumerate}

\citet{koll2015deciphering} modelled cool tidally locked planets and reported that the phase curve amplitude (i.e. the day-night contrast) depends mainly on the ratio of advective to radiative timescales, along the lines sketched out above.  \citet{zhang2016effects} introduced a more general quantitative estimate of the day-night contrast which takes Coriolis effects into account. It is based on simplified solutions of the primitive equations from \citet{komacek2016atmospheric}.  In the regime where surface friction is not the controlling factor and the Coriolis effects are comparable to or dominate advection effects, their result reads
\begin{equation}\label{eq:dn_contrast}
 \frac{\Delta T}{\Delta T_{eq}} \sim 1 - 1 / (1 +\frac{t_{wave}^{2} \Omega}{t_{rad}\Delta \ln p})
\end{equation}
where $\Delta T_{eq}$ is the radiative-equilibrium day-night contrast (i.e. without any convection or bulk dynamics), and $\Delta \ln p$ is the difference in log pressure between the surface and the relevant pressure level (see \citet{zhang2016effects} for details).

Eq. \ref{eq:dn_contrast} can be recast in the form
\begin{equation}\label{eq:dn_contrast1}
 \frac{\Delta T}{\Delta T_{eq}} \sim 1 - 1 / (1 + \Lambda^{-1}\frac{t_{wave}}{t_{rad}\Delta \ln p})
\end{equation}
which highlights the influence of the WTG parameter $\Lambda$. One can make the day-night temperature contrast weak either by making the WTG parameter large or the nondimensional radiative damping time large.

To estimate the hot spot shift, \citet{zhang2016effects} (see their Eq. 47)  introduce a kinematic theory balancing radiative damping against advection by a specified jet speed $U$.  The hot spot shift is a monotonically increasing function of $t_{rad}/t_{adv}$, with the hotspot located at the substellar point for $t_{rad}/t_{adv} \ll 1$ and moving to the terminator as $t_{rad}/t_{adv} \rightarrow \infty$.

The atmospheric circulation determining the hot-spot shift and phase curve amplitude is governed by $\Lambda$ and the nondimensional radiative damping time (e.g. $t_{rad}/t_{wave}$). the latter depends on surface pressure in the combination $p_s/\mu$ whereas the former depends on $\mu$ alone. It is this property that opens the possibility of estimating both $p_s$ and $\mu$ through interpretation of the observed phase curve. The results quoted from \citet{zhang2016effects} suggest that the hot-spot phase shift depends on the nondimensional $t_{rad}$ alone, whereas the phase curve amplitude depends on $\Lambda$ as well.

Figure \ref{fig:param_map} shows  the parameter space we are investigating, and the regimes with significant hot-spot shift or day-night contrast, calculated using equation \ref{eq:dn_contrast} and Eq. 47 of \citet{zhang2016effects}. According to these formulae the area to the right of the blue line has a significant day-night contrast (80\% of the equilibrium value with no heat transport), while the area to the left of the red line has a significant (\textgreater20$\degr$) hot-spot shift. An atmosphere in the green area between these lines has the best chance to replicate the observed phase curve. The regime boundaries shown in the figure serve only to help situate our simulations in parameter space; our interpretation of the observed phase curve is based on GCM simulations, which depend on the same fundamental nondimensional parameters but take into account numerous physical effects, such as changes in jet width and strength, not captured by the simple scaling laws.

The infrared optical thickness $\tau_{\infty}$ is a third nondimensional parameter of the climate.  In the configuration we model, it is of secondary importance so far as the character of the circulation is concerned, since in a mostly convective atmosphere it only sets the surface temperature relative to the radiating temperature. Increasing $\tau_{\infty}$ somewhat increases the wave speed through increasing the temperature, but not significantly so over the range of $\tau_{\infty}$ we consider.  In the case where the atmosphere has a window region near the wavelength at which observations are taken, however, so that the observed brightness temperature reflects temperatures deeper in the atmosphere than the grey-gas radiating level, increasing $\tau_{\infty}$ will significantly increase the amplitude of the phase curve.

 %%%%%%%%%%%%%%%%%%%%%%%%%%%%%%%%%%%%%%%%%%%%%%%%%%%%%%%%%%%%%%%%%%%%%%%%

\section{Results}\label{sec:results}

\begin{deluxetable}{cccccc}
\tablenum{1}
\tablecaption{Exo-FMS Test Parameters. The first three tests are intended to show the different regimes discussed in section \ref{sec:theory}, and Test 5 is the ``best-fit'' composition discussed in section \ref{sec:ps_effect}. Note that the molar heat capacity of H\textsubscript{2} and N\textsubscript{2} is almost the same. The opacity $\kappa$ has been calculated from the chosen $\tau_{\infty}$. The 10 bar H\textsubscript{2} opacity is consistent with the CIA-only results of \citet{freedman2014cia}.}\label{tab:paramstable}
\tablehead{
Test &  \colhead{p\textsubscript{s}}  & \colhead{$\mu$} & \colhead{$\tau_{\infty}$} & \colhead{$\kappa$} \\
&  \colhead{(bar)} & \colhead{(gmol\textsuperscript{-1})} & & \colhead{(cm\textsuperscript{2}kg\textsuperscript{-1})}
}
\startdata
1 -- H\textsubscript{2} & 10 & 2.0 & 8.0 & 22.4  \\
2 -- N\textsubscript{2} & 10 & 28.0 & 8.0 & 22.4 \\
\\
3 -- H\textsubscript{2}+N\textsubscript{2} & 10 & 4.6 & 8.0 & 22.4 \\
4 -- H\textsubscript{2}+N\textsubscript{2} & 5 & 4.6 & 4.0 & 22.4 \\
5 -- H\textsubscript{2}+N\textsubscript{2} & 3 & 4.6 & 2.4 & 22.4 \\
\\
6 -- H\textsubscript{2}+N\textsubscript{2} & 10 & 15.0 & 8.0 & 22.4 \\
7 -- H\textsubscript{2}+N\textsubscript{2} & 5 & 15.0 & 4.0 & 22.4 \\
8 -- H\textsubscript{2}+N\textsubscript{2} & 3 & 15.0 & 2.4 & 22.4 \\
\\
9 -- H\textsubscript{2}+N\textsubscript{2} & 5 & 4.6 & 2.0 & 11.2 \\
10 -- H\textsubscript{2}+N\textsubscript{2} & 5 & 4.6 & 8.0 & 44.8 \\
\enddata

\end{deluxetable}

This section shows the results of our tests in Exo-FMS. We modelled ten different atmospheres and compared their phase curves and temperature distributions to those measured and reconstructed by \citet{demory2016map}. Sections \ref{sec:model_results}, \ref{sec:ps_effect}, and \ref{sec:tauinf_effect} investigate the effects of mean molecular weight $\mu$, surface pressure p\textsubscript{s}, and optical thickness $\tau_{inf}$. We focus on the hot-spot shift and day-night contrast of the temperature distributions, which correspond to the offset and amplitude of the phase curves. We then discuss which atmospheres are consistent with the observed phase curve, and how real-gas radiative transfer and cloud formation could fully explain the observations. %***Replace "six" here with the current number

\begin{deluxetable}{ccccc}
\tablenum{2}
\tablecaption{Exo-FMS Results Summary. Hot-spot shift and day-night contrast are taken from the half-surface-pressure level, which does not display the largest shift and contrast but is useful for comparison. The day-night contrast is the difference between the warmest and coldest hemispheres, for consistency with \citet{demory2016map}. These results are a summary of the broad features of each test, and Section \ref{sec:results} discusses these features and the reconstructed phase curves in much greater detail.}\label{tab:resultstable}
\tablehead{
\colhead{Test} & \colhead{p\textsubscript{s}} & \colhead{$\mu$} & \colhead{Hot-spot} & \colhead{Day-night} \\
&  \colhead{(bar)} & \colhead{(gmol\textsuperscript{-1})} & & \colhead{(K)}
}
\startdata
Observations & &  & +(41 $\pm$ 12) & (1300 $\pm$ 670)   \\
\\
1 -- H\textsubscript{2} & 10 & 2.0 & +45$\degr$ & 100  \\
2 -- N\textsubscript{2} & 10 & 28.0 & 0$\degr$ & 750 \\
\\
3 -- H\textsubscript{2}+N\textsubscript{2} & 10 & 4.6 & +30$\degr$ & 200 \\
4 -- H\textsubscript{2}+N\textsubscript{2} & 5 & 4.6 & +25$\degr$ & 250 \\
5-- H\textsubscript{2}+N\textsubscript{2} & 3 & 4.6 & +15$\degr$ & 300 \\
\\
6 -- H\textsubscript{2}+N\textsubscript{2} & 10 & 15.0 & 0$\degr$ & 150 \\
7-- H\textsubscript{2}+N\textsubscript{2} & 5 & 15.0 & 0$\degr$ & 550 \\
8 -- H\textsubscript{2}+N\textsubscript{2} & 3 & 15.0 & 0$\degr$ & 600 \\
\\
9 -- H\textsubscript{2}+N\textsubscript{2} & 5 & 4.6 & +20$\degr$ & 200 \\
10 -- H\textsubscript{2}+N\textsubscript{2} & 5 & 4.6 & +25$\degr$ & 250 \\
\enddata
\end{deluxetable}

\subsection{Effect of Mean Molecular Weight}\label{sec:model_results}

\begin{figure*}
 \gridline{\fig{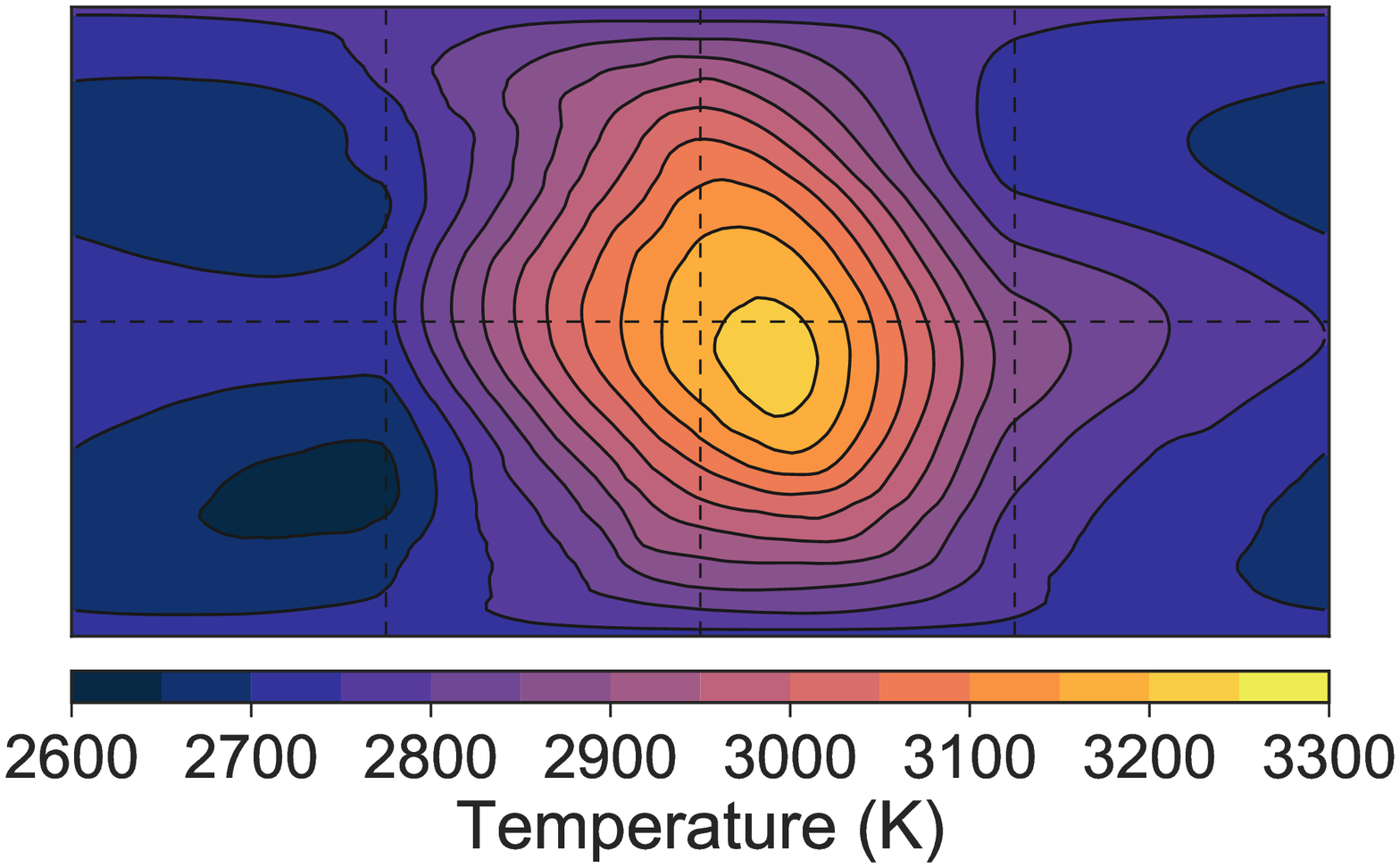}{0.3\textwidth}{H\textsubscript{2}, 10 bar: surface air temperature}
           \fig{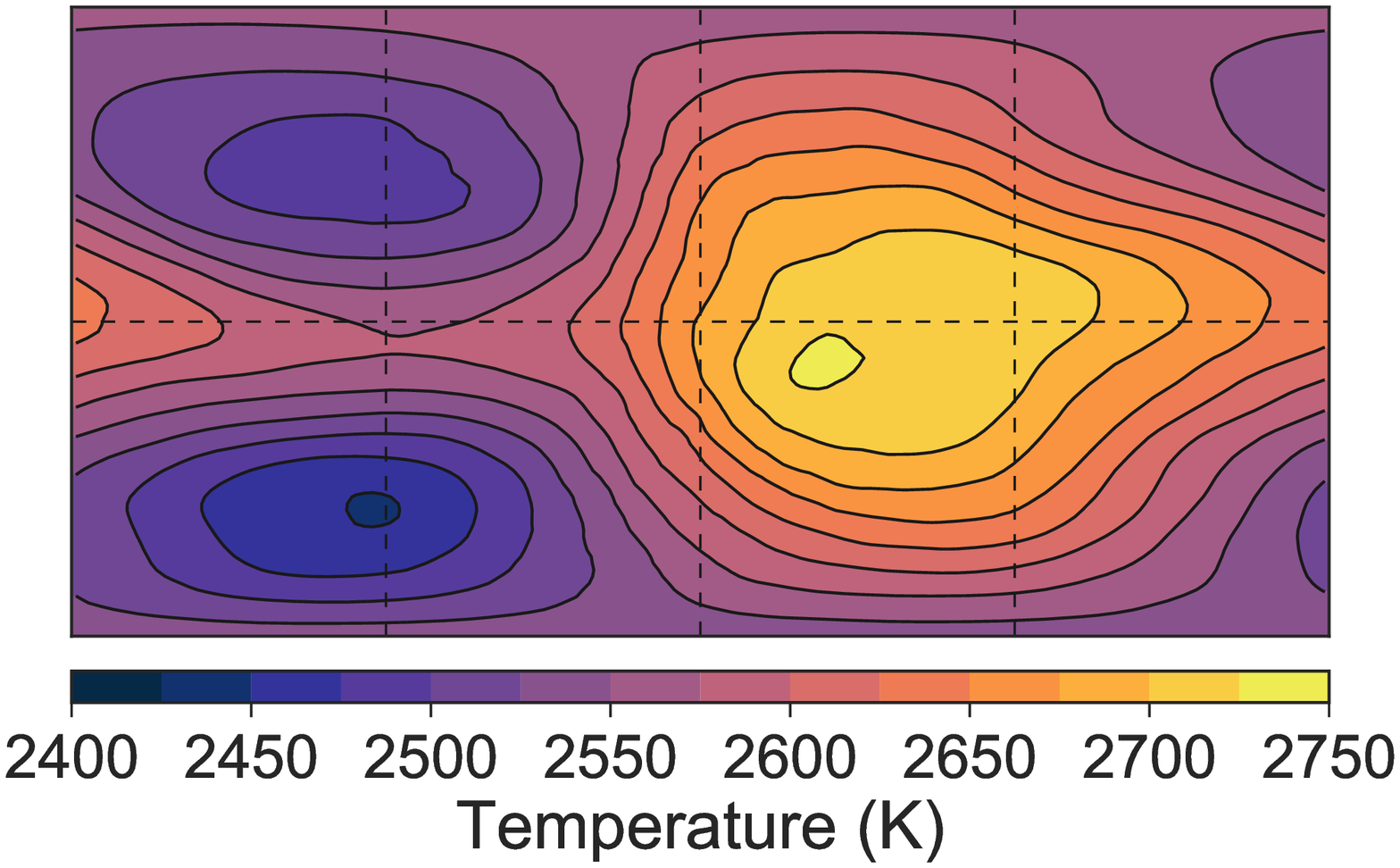}{0.3\textwidth}{H\textsubscript{2}, 10 bar: half-surface-pressure air temperature}
           \fig{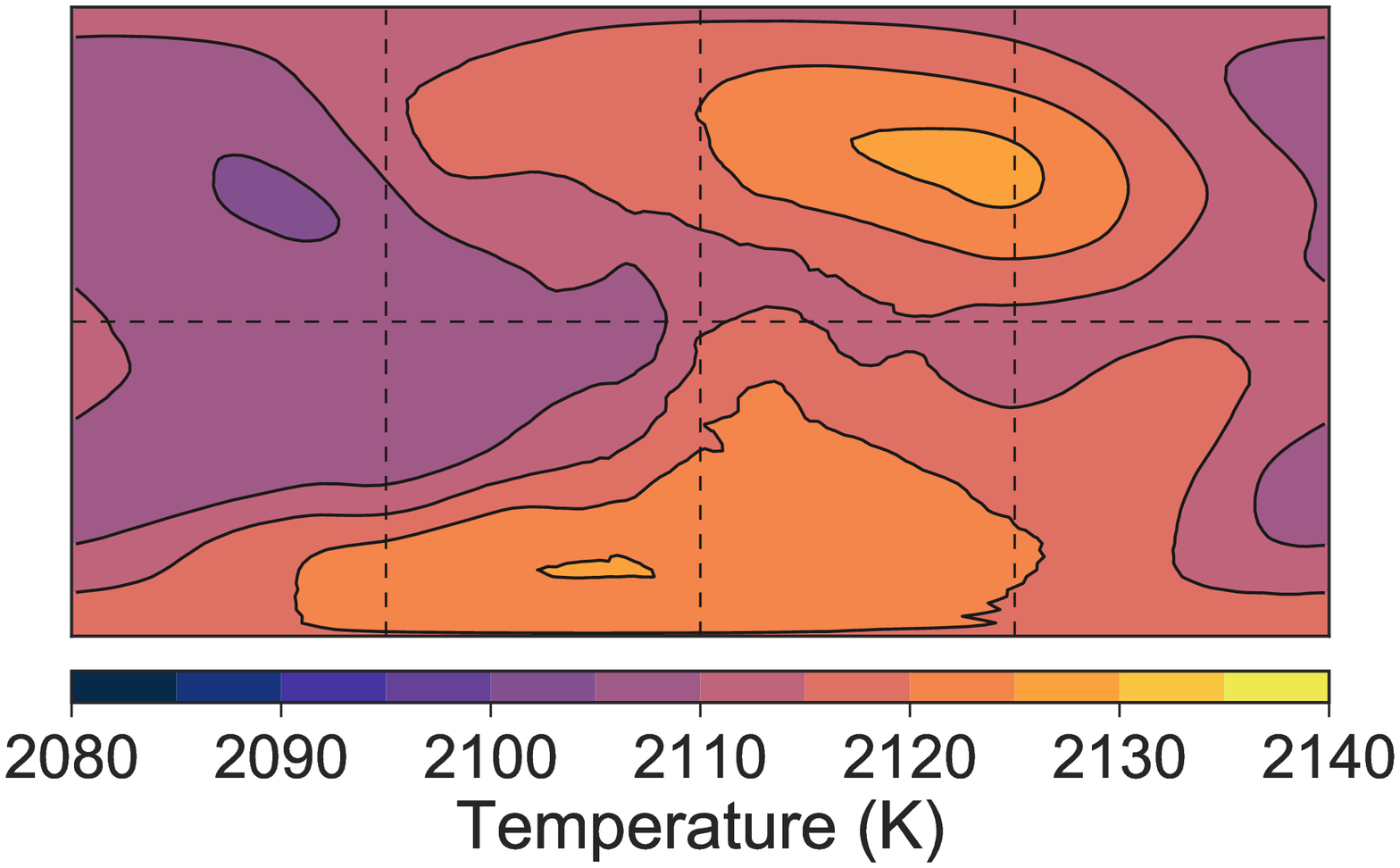}{0.3\textwidth}{H\textsubscript{2}, 10 bar: brightness temperature}}
 \gridline{\fig{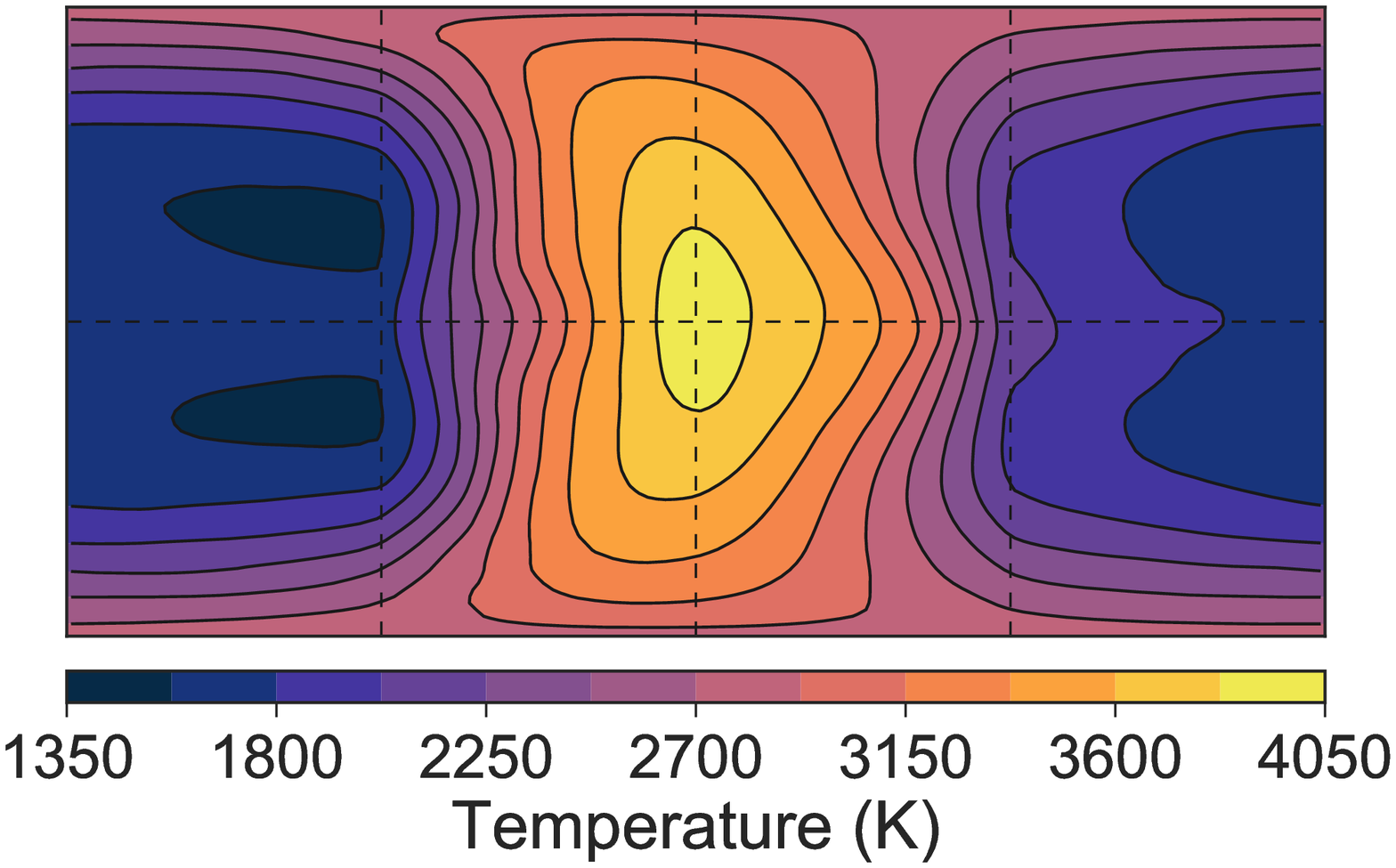}{0.3\textwidth}{N\textsubscript{2}, 10 bar: surface air temperature}
           \fig{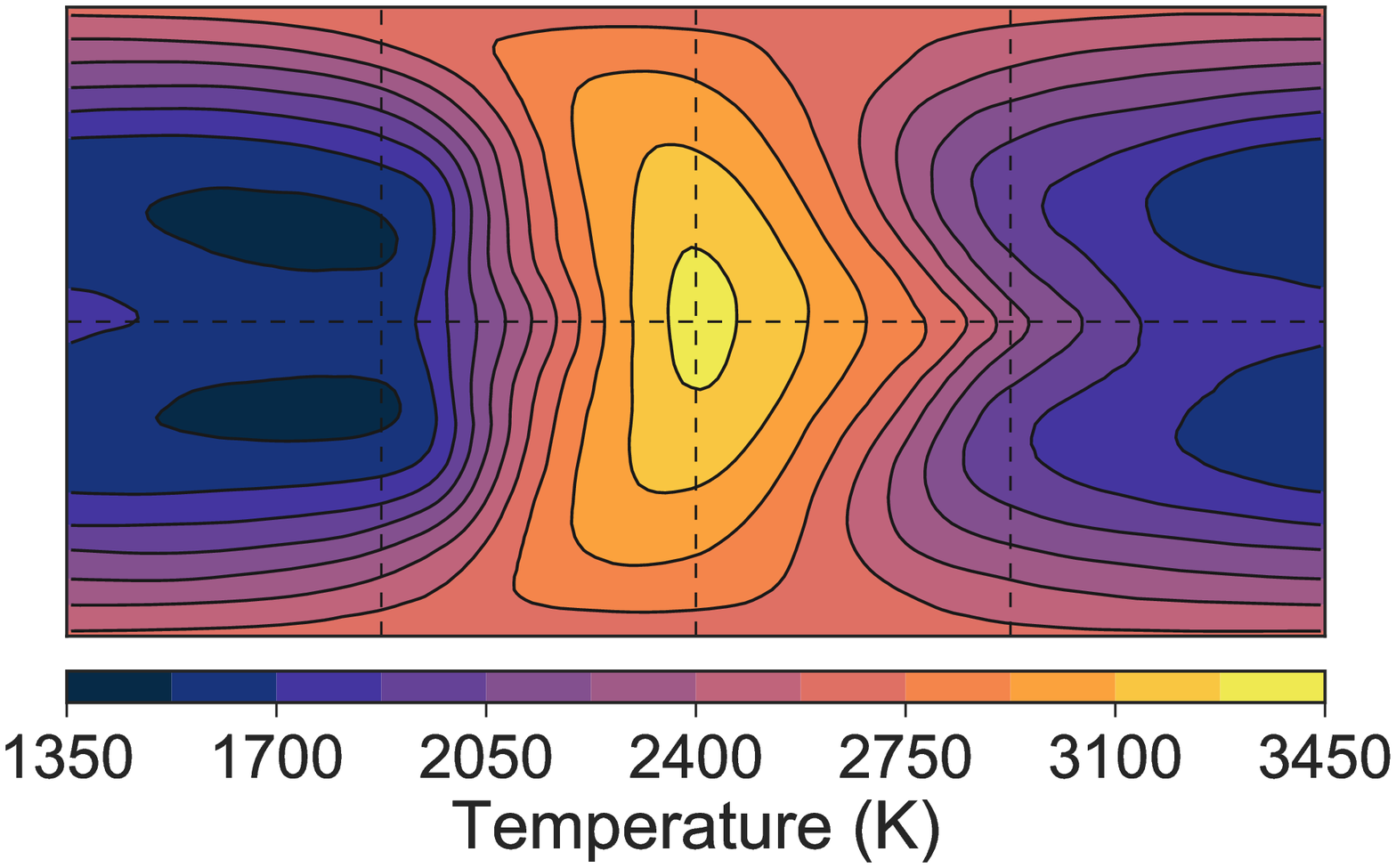}{0.3\textwidth}{N\textsubscript{2}, 10 bar: half-surface-pressure air temperature}
           \fig{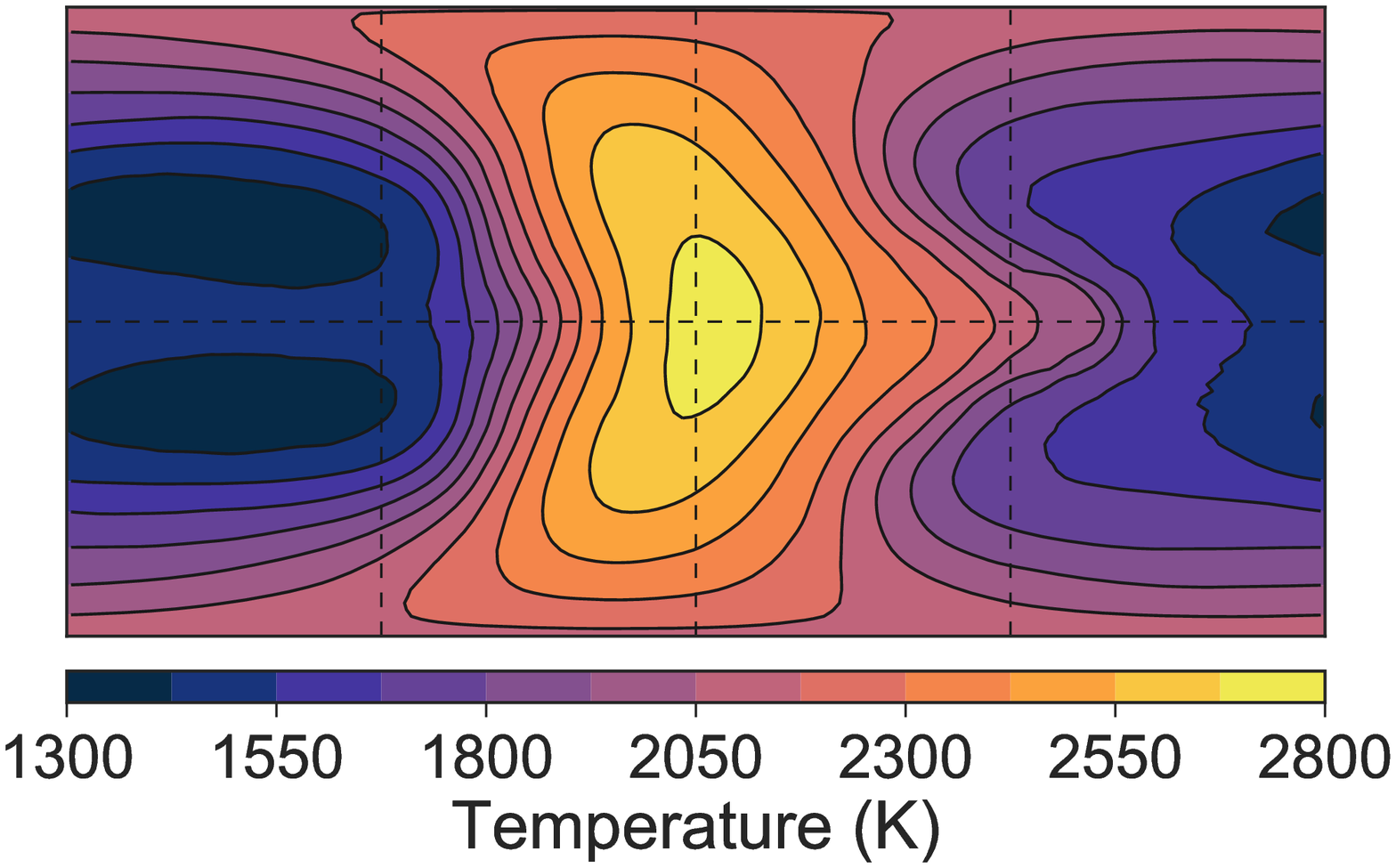}{0.3\textwidth}{N\textsubscript{2}, 10 bar: brightness temperature}}
 \gridline{\fig{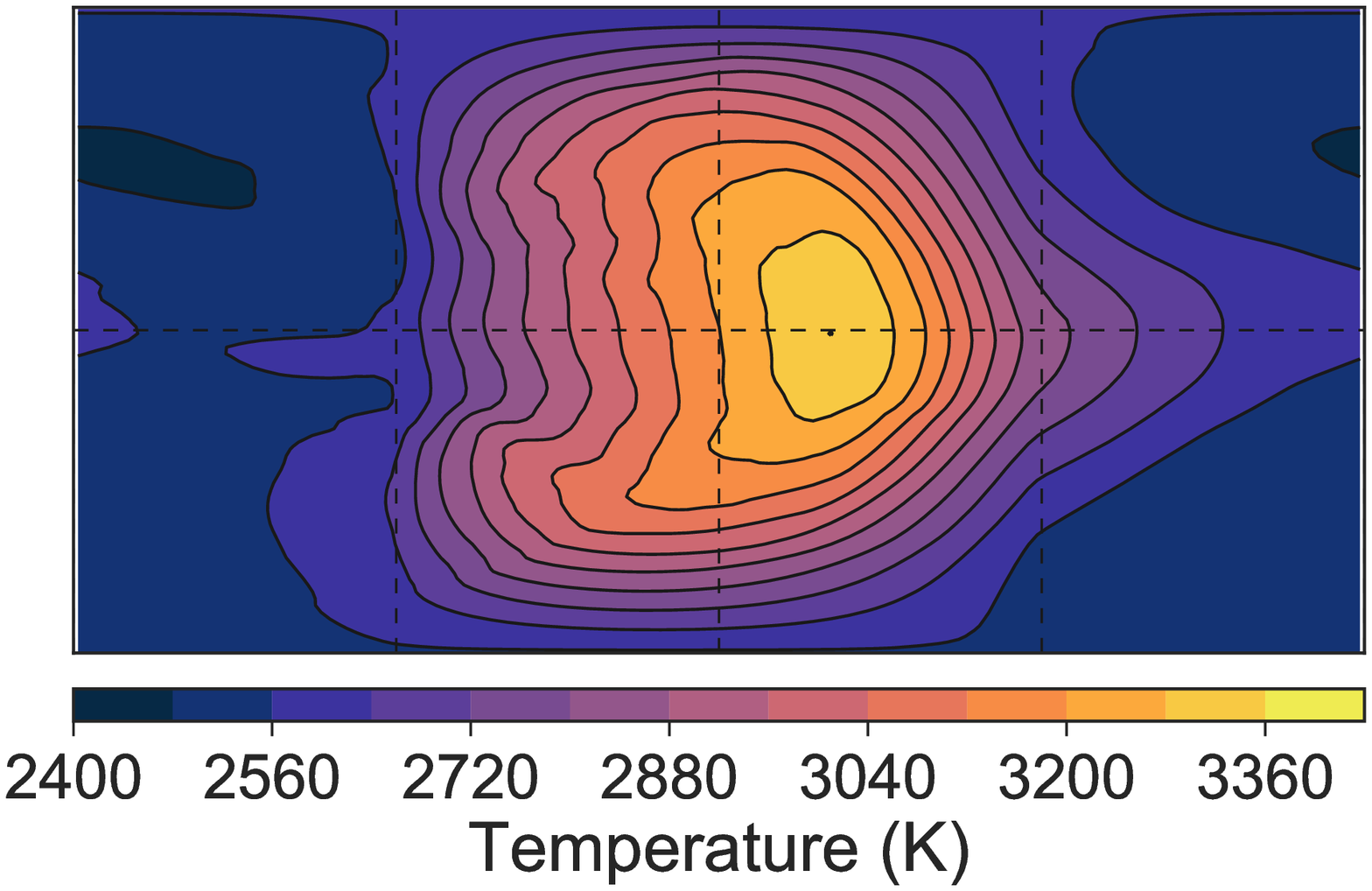}{0.3\textwidth}{H\textsubscript{2}+N\textsubscript{2}, $4.6\ \mathrm{gmol}^{-1}$, 10 bar: surface air temperature}
           \fig{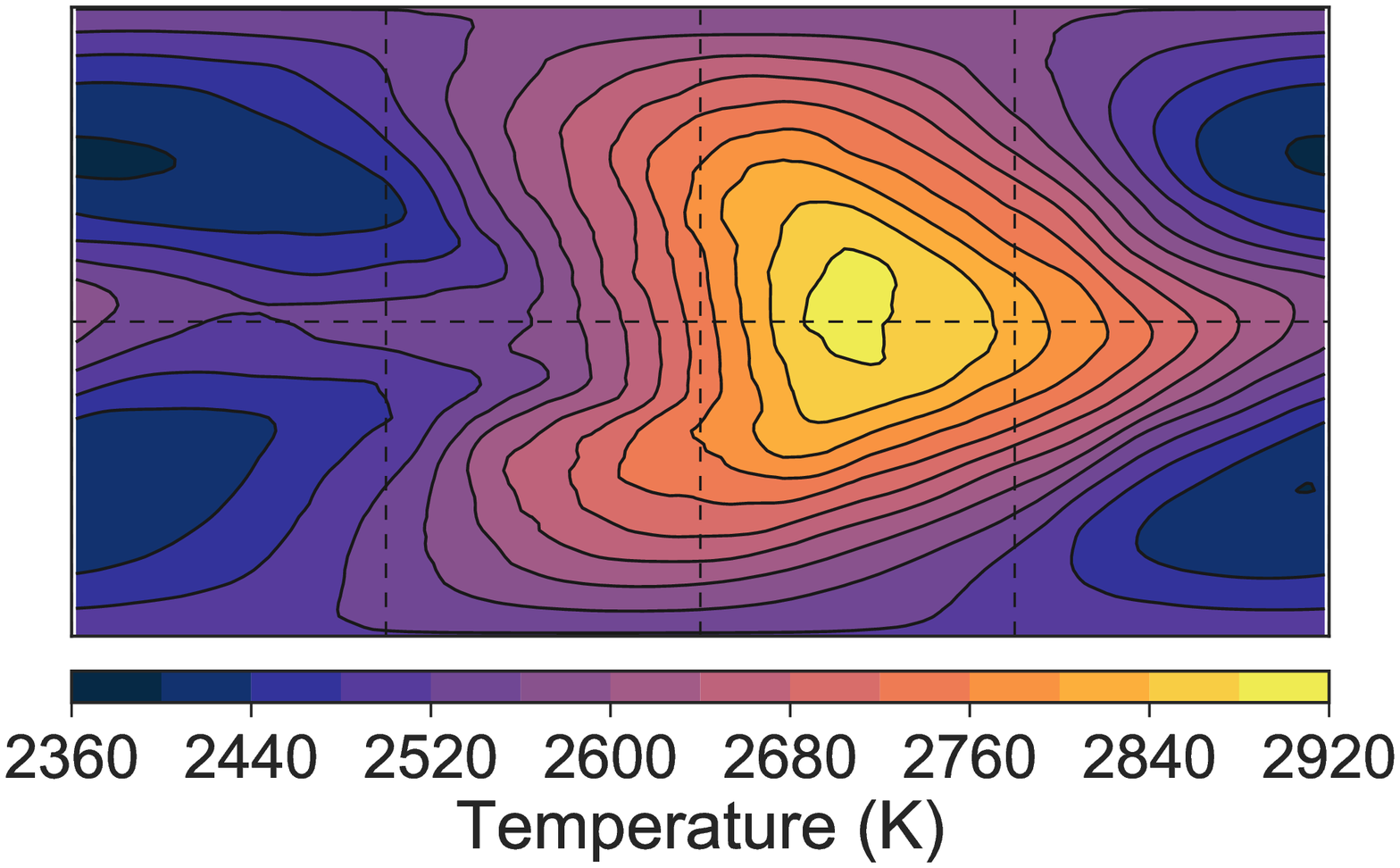}{0.3\textwidth}{H\textsubscript{2}+N\textsubscript{2}, $4.6\ \mathrm{gmol}^{-1}$, 10 bar: half-surface-pressure air temperature}
           \fig{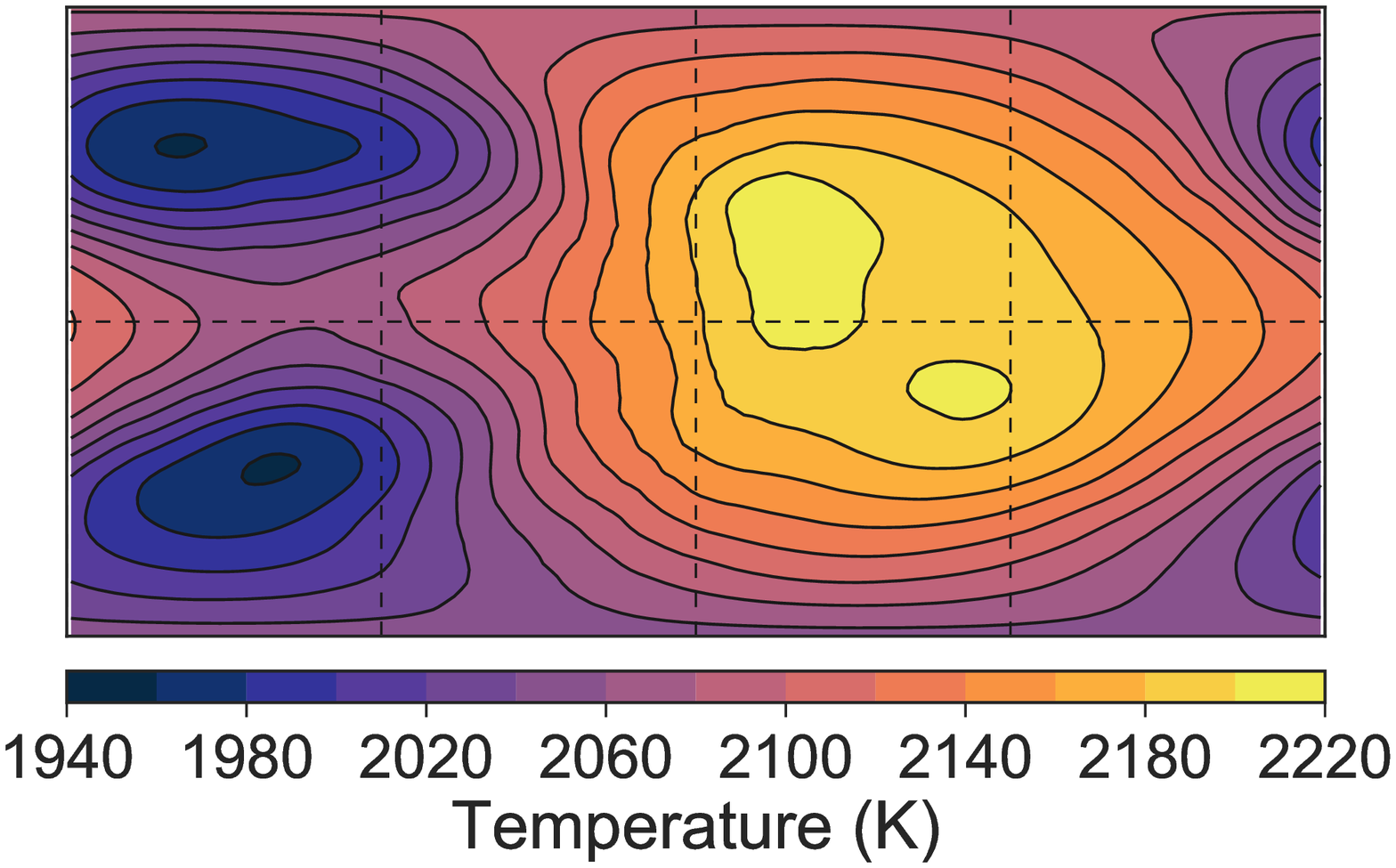}{0.3\textwidth}{H\textsubscript{2}+N\textsubscript{2}, , $4.6\ \mathrm{gmol}^{-1}$, 10 bar: brightness temperature}}
\caption{10-day average temperature maps centred on the substellar point, after each Exo-FMS test reaches top-of-atmosphere radiative balance. The first column is surface air temperature, which generally shows the largest day-night contrast as it is strongly coupled to the surface temperature and incoming stellar flux. The second column is the half-pressure air temperature, which generally shows the largest hot-spot shift as it is normally close to the height of the superrotating jet. The third column is the brightness temperature, which determines the thermal phase curve for the model grey-gas radiation -- in reality, the thermal emission could be from a different level of the atmosphere.\label{fig:allT_results}}
\end{figure*}

We use a pure H\textsubscript{2} atmosphere as a starting point, as the measurements of  \citet{tsiaras2016detection} suggested that the atmosphere is H\textsubscript{2}-rich.

Test 1 is a pure H\textsubscript{2} atmosphere, with surface pressure $p_{s} = 10\ \mathrm{bar}$ and optical thickness $\tau_{\infty}= 8.0$; for this case, the surface temperature and vertical structure of the atmosphere are very similar to results that would be obtained with a real-gas radiative transfer calculation taking into account the collisional opacity of $\mathrm{H_2}$. The theory in Section \ref{sec:theory} predicts that this test will have a large hot-spot shift but small day-night contrast, because its radiative timescale is much longer than its transport timescale. The results confirm this, showing a weak day-night temperature gradient but significant hot-spot shift at both the surface and mid-atmosphere levels, though with a more pronounced shift at mid-atmosphere than at the surface. The vertical structure of the temperature pattern will be discussed in Section \ref{sec:vertical_structure}.  The grey brightness temperature $T_{b}$ shown in the third column provides a direct indication of the net horizontal heat transport, as the infrared cooling to space is $\sigma T_{b}^{4}$. Very weak heat transport would manifest as a close resemblance of $\sigma T_{b}^{4}$ to the instellation pattern, whereas complete horizontal temperature homogenization manifests as a uniform $T_{b}$. The latter is very nearly the case for this atmosphere, to the extent that the variations in $T_{b}$ are so small that a statistically stationary pattern has not yet fully emerged. In any case, the temperature is too uniform at all levels to be compatible with the observed phase curve.
Low molecular weight atmospheres strongly favour weak temperature gradients because the WTG parameter $\Lambda$ and the radiative damping time $t_{rad}$ both become larger as molecular weight is decreased.

Test 2 is a pure N\textsubscript{2} atmosphere, with surface pressure $p_{s} = 10\ \mathrm{bar}$ and optical thickness held fixed at $\tau_{\infty}= 8.0$. This tested the effect of changing the molecular weight of the atmosphere. Section \ref{sec:theory} predicts that N\textsubscript{2} atmosphere will have a large day-night contrast but a small hot-spot shift, as its radiative timescale is much shorter than its transport timescale, and the WTG parameter is also smaller.  The results confirm this, as the day-night temperature gradient is large but the hot-spot shift is very small at all levels. Section \ref{sec:sim_pc} shows how this results in a phase curve with a large amplitude but little to no peak offset. The comparison of Test 1 to Test 2 is consistent with the simulations of \citet{kataria2014atmospheric} showing that low molecular weight favours an increased phase shift.

We used these results and the theory in Section \ref{sec:theory} to select a composition which might match the observations. Figure \ref{fig:param_map} shows the predicted regimes in our parameter space. The shaded area is the region which should support both a hot-spot shift and day-night contrast. Our first test in this region was Test 3 -- an H\textsubscript{2}-N\textsubscript{2} mixture, with mean molecular weight $4.6\ \mathrm{gmol}^{-1}$, surface pressure $p_{s} = 10\ \mathrm{bar}$ and optical thickness $\tau_{\infty}= 8.0$. Section \ref{sec:theory} predicts that this composition could support a large hot-spot shift and a large day-night contrast. Figure \ref{fig:allT_results} and Table \ref{tab:resultstable} show that both the brightness temperature and the temperature at the half-pressure level have a large hot-spot and day-night contrast, although not as large at the observations. In the rest of this paper, we investigate the effect of surface pressure and optical thickness on the temperature distribution and thermal phase curve, to test our expectations in Section \ref{sec:theory} and to find an atmospheric composition which better matches the observations.

In summary, Tests 1, 2 and 3 do not match the measured temperature distribution, but do confirm the general effects of changing the atmospheric properties described in section \ref{sec:theory}. Table \ref{tab:paramstable} lists the test cases and their parameters, and Table \ref{tab:resultstable} summarises the results.

%%%%%%%%%%%%%%%%%%%%%%%%%%%%%%%%%%%%%%%%%%%%%%%%%%%

\subsection{Effect of Surface Pressure}\label{sec:ps_effect}

\begin{figure*}[t]
 \gridline{\fig{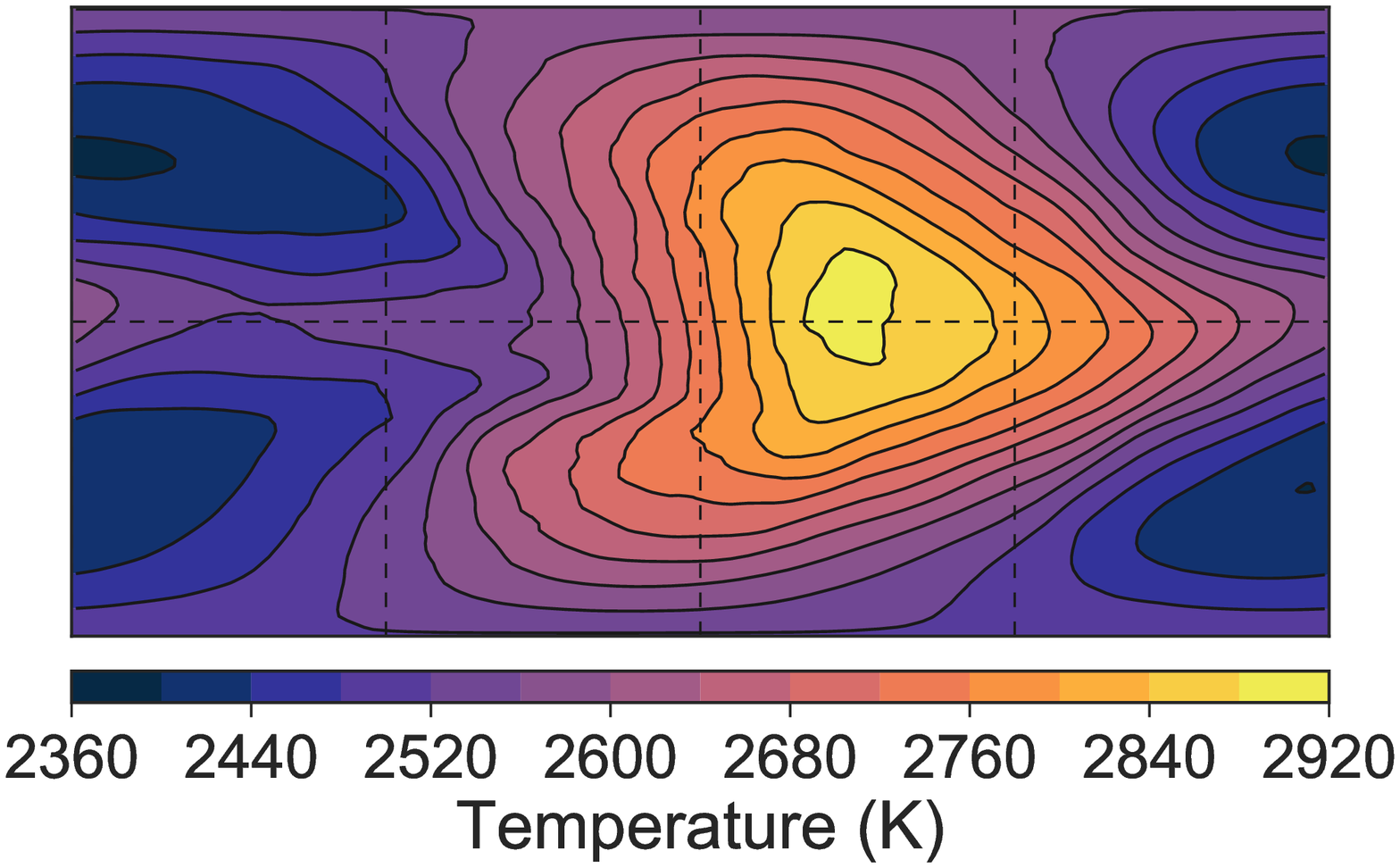}{0.3\textwidth}{$4.6\ \mathrm{gmol}^{-1}$, H\textsubscript{2}+N\textsubscript{2}, 10 bar: half-surface-pressure air temperature}
           \fig{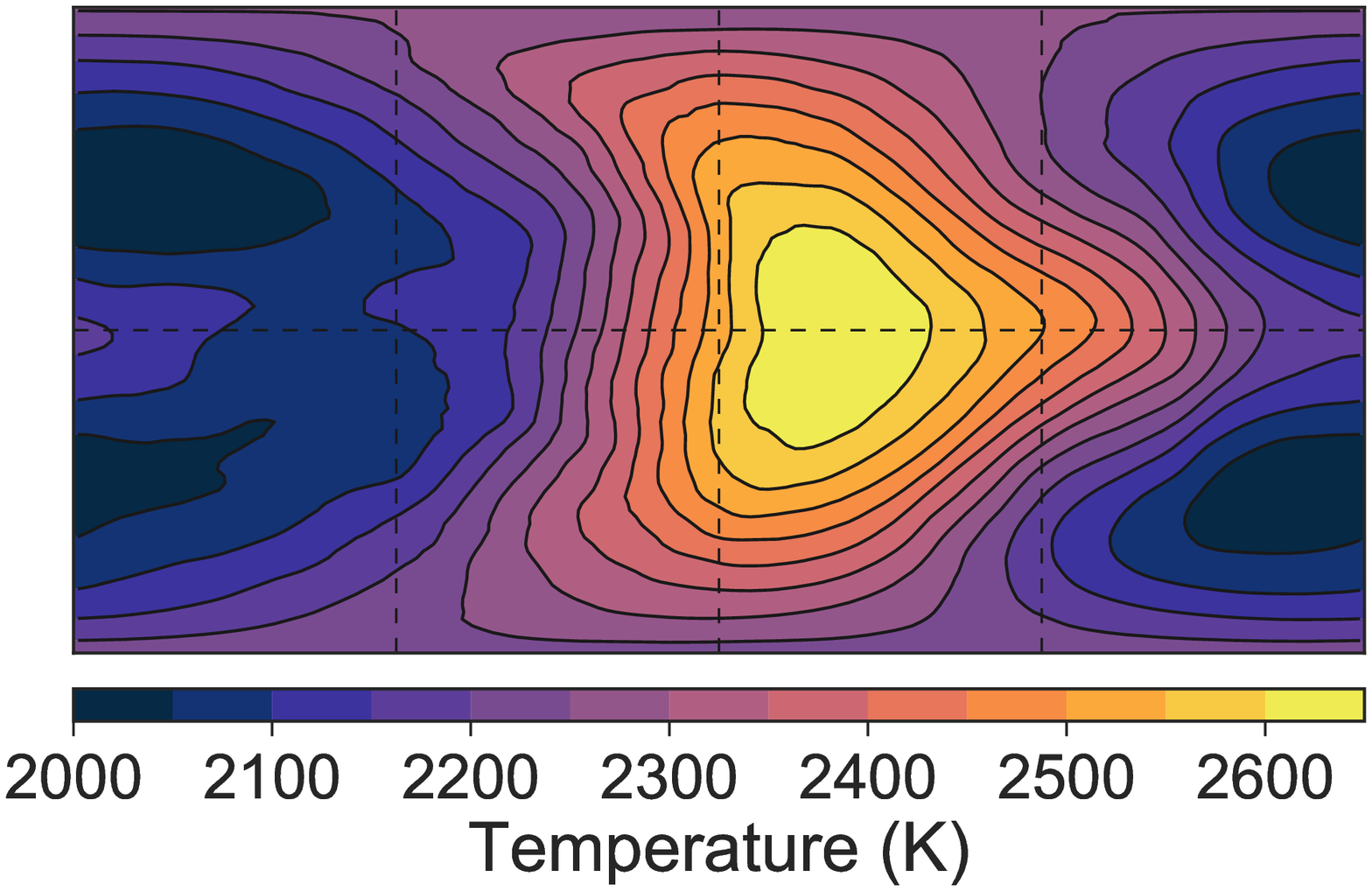}{0.3\textwidth}{$4.6\ \mathrm{gmol}^{-1}$, H\textsubscript{2}+N\textsubscript{2}, 5 bar: half-surface-pressure air temperature}
           \fig{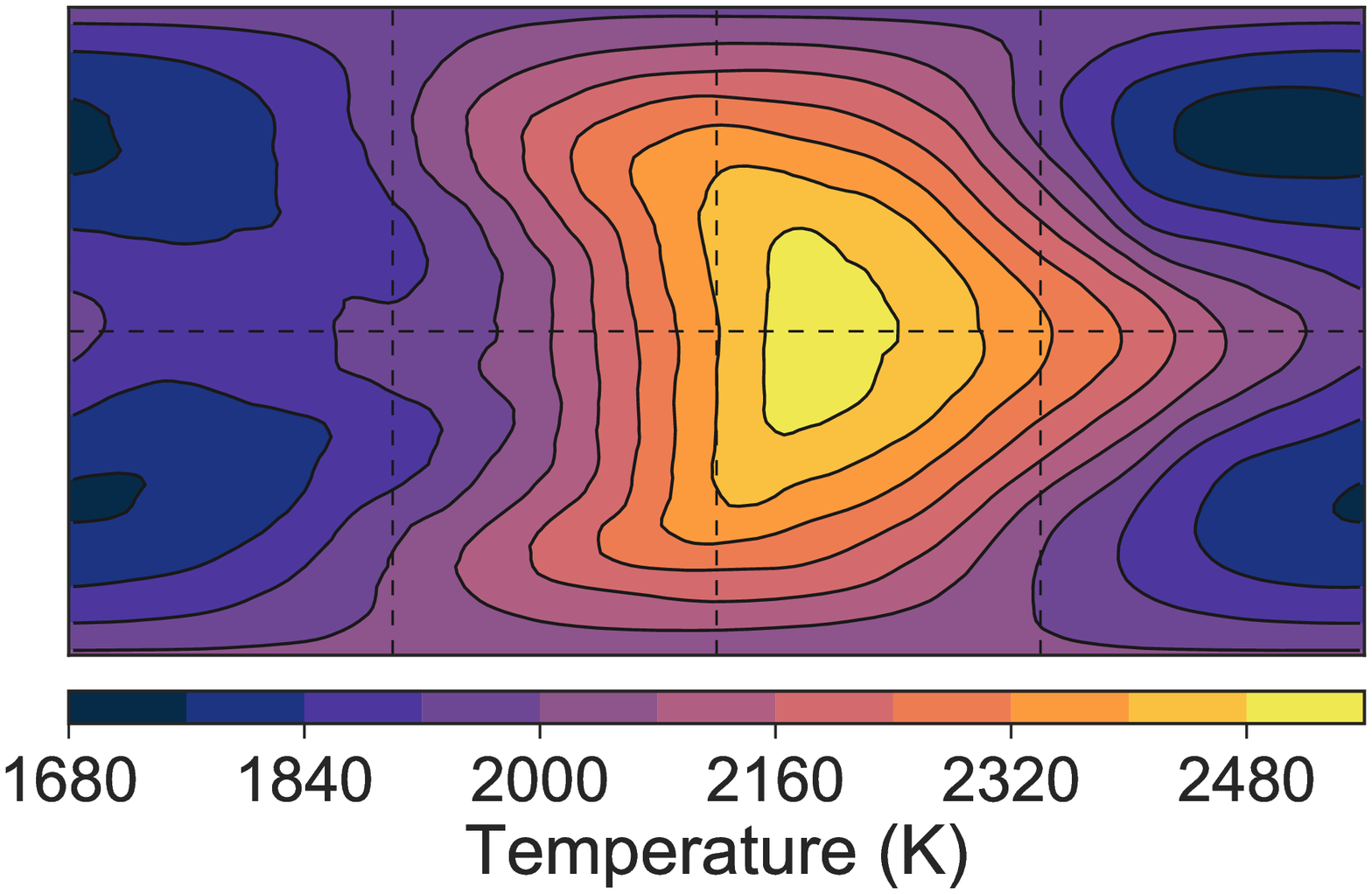}{0.3\textwidth}{$4.6\ \mathrm{gmol}^{-1}$, H\textsubscript{2}+N\textsubscript{2}, 3 bar: half-surface-pressure air temperature}}

 \gridline{\fig{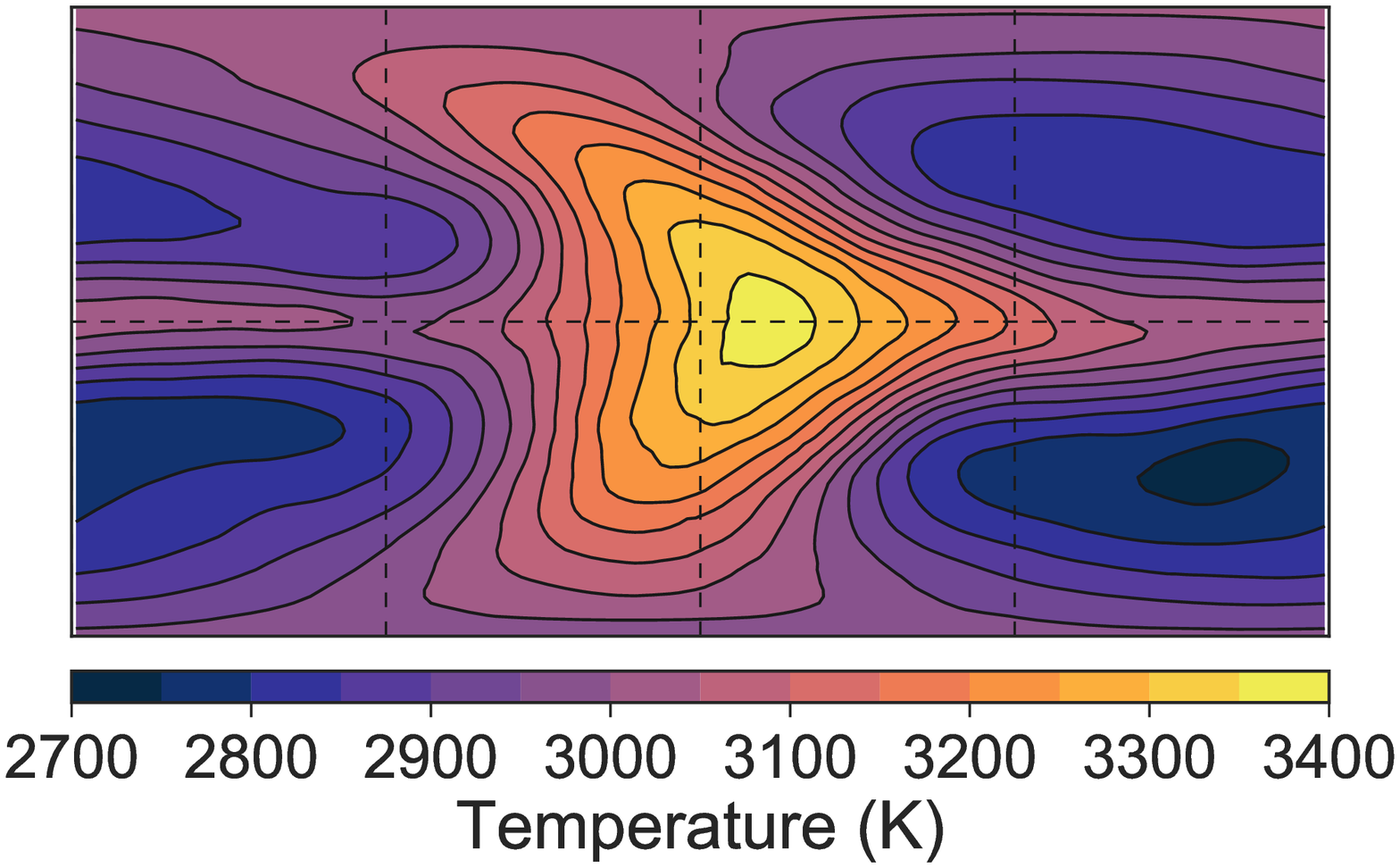}{0.3\textwidth}{$15\ \mathrm{gmol}^{-1}$, H\textsubscript{2}+N\textsubscript{2}, 10 bar: half-surface-pressure air temperature}
           \fig{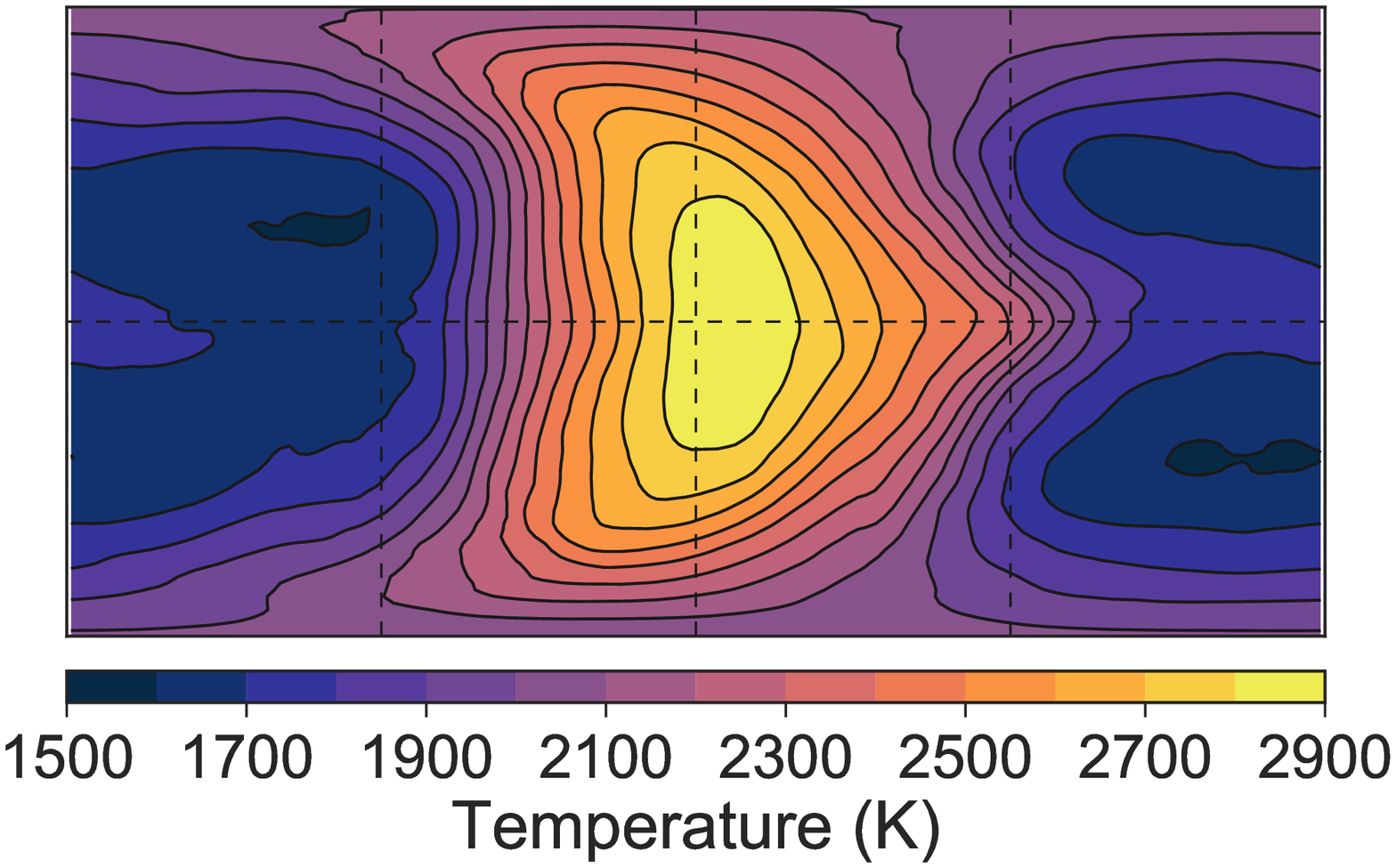}{0.3\textwidth}{$15\ \mathrm{gmol}^{-1}$, \textsubscript{2}+N\textsubscript{2}, 5 bar: half-surface-pressure air temperature}
           \fig{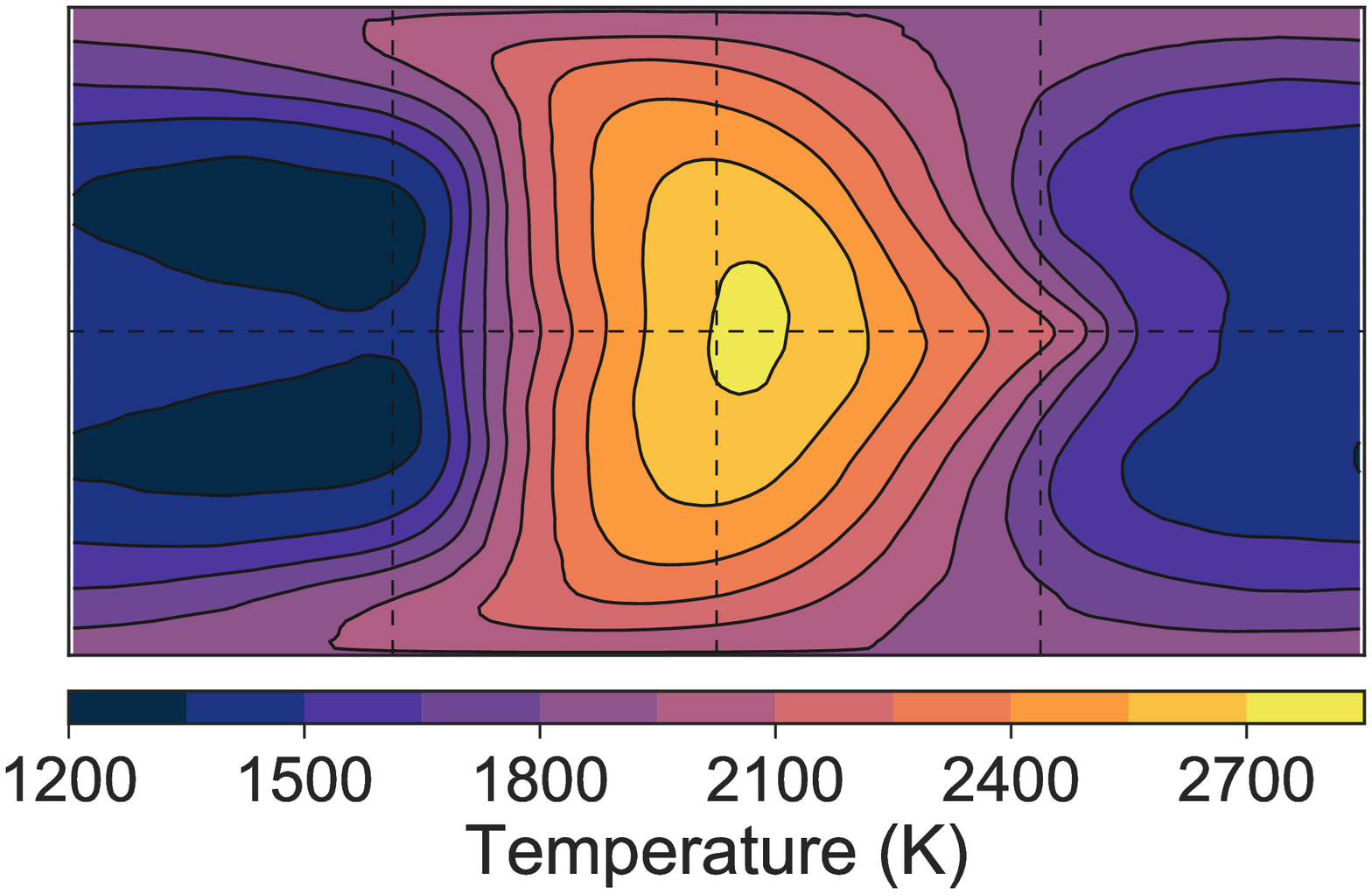}{0.3\textwidth}{$15\ \mathrm{gmol}^{-1}$, 15H\textsubscript{2}+N\textsubscript{2}, 3 bar: half-surface-pressure air temperature}}

\caption{10-day average temperature maps of the half-surface-pressure level for $4.6\ \mathrm{gmol}^{-1}$ and $15\ \mathrm{gmol}^{-1}$ H\textsubscript{2} + N\textsubscript{2} atmospheres with surface pressures of 3, 5, and 10 bar, and constant opacity $\kappa = 22.4$ cm\textsuperscript{2}kg\textsuperscript{-1}. The global temperature distributions are similar but the maximum and minimum temperatures differ.\label{fig:H2N2_T_maps}}
\end{figure*}

In this section we discuss a number of simulations with surface pressures of 3, 5, and 10 bar. The tests either have mean molecular weight $\mu = 4.6\ \mathrm{gmol}^{-1}$ or $15\ \mathrm{gmol}^{-1}$, corresponding to an H\textsubscript{2}-N\textsubscript{2} mixture with molar concentrations of 90\%  H\textsubscript{2} and 10\%   N\textsubscript{2}, or 50\%  H\textsubscript{2} and 50\%   N\textsubscript{2}. We used the same opacity $\kappa$ in all these tests.

We will focus on the temperature of the half-pressure level, which fits the observations better in general (see Section \ref{sec:vertical_structure}). The observed high maximum brightness temperature suggests that the greenhouse gas supplying the optical thickness has a window at $4.5\ \micron$, as an optically thick grey-gas-like continuum would have a radiating level high in the atmosphere, which does not fit the observed high day-side temperature and large day-night contrast.

Figure \ref{fig:H2N2_T_maps} shows the temperature at the half-surface-pressure level in the tests with different surface pressures. Increasing the atmospheric pressure affects the temperature distribution as predicted in Section \ref{sec:theory}, as the day-night contrast decreases and the hot-spot shift increases (see Table \ref{tab:resultstable}). The hot-spot shift does not increase for the $15\ \mathrm{gmol}^{-1}$, which may be because their higher mean molecular weight forces a very short radiative timescale even at higher surface pressures.

The $4.6\ \mathrm{gmol}^{-1}$, 10 bar case is compatible with the observed phase curve maximum and shift (see Section \ref{sec:sim_pc} for a more quantitative comparison), but has a much hotter night-side than the observations. The 5 bar case matched the observed peak shift and amplitude within error, and has a cooler night-side which is more compatible with observations (see Section \ref{sec:condensables} for a discussion of how clouds could further improve the night-side fit). Therefore, we chose the $4.6\ \mathrm{gmol}^{-1}$, 5 bar case as our ``best-fit'' test.

In the rest of this paper, we discuss the effect of vertical structure on the brightness temperature and phase curve. We simulate the phase curves of our tests and compare them to the observed phase curve. We will also consider which other physical processes such as cloud formation could affect the real temperatures and observed fluxes to explain the observed phase curve.

%%%%%%%%%%%%%%%%%%%%%%%%%%%%%%%%%%%%%%%%%%%%%%%%%%%%%%%%%%%%%%%%%%%%%%%%%%

\subsection{Effect of Optical Thickness}\label{sec:tauinf_effect}

\begin{figure*}[t]
 \gridline{\fig{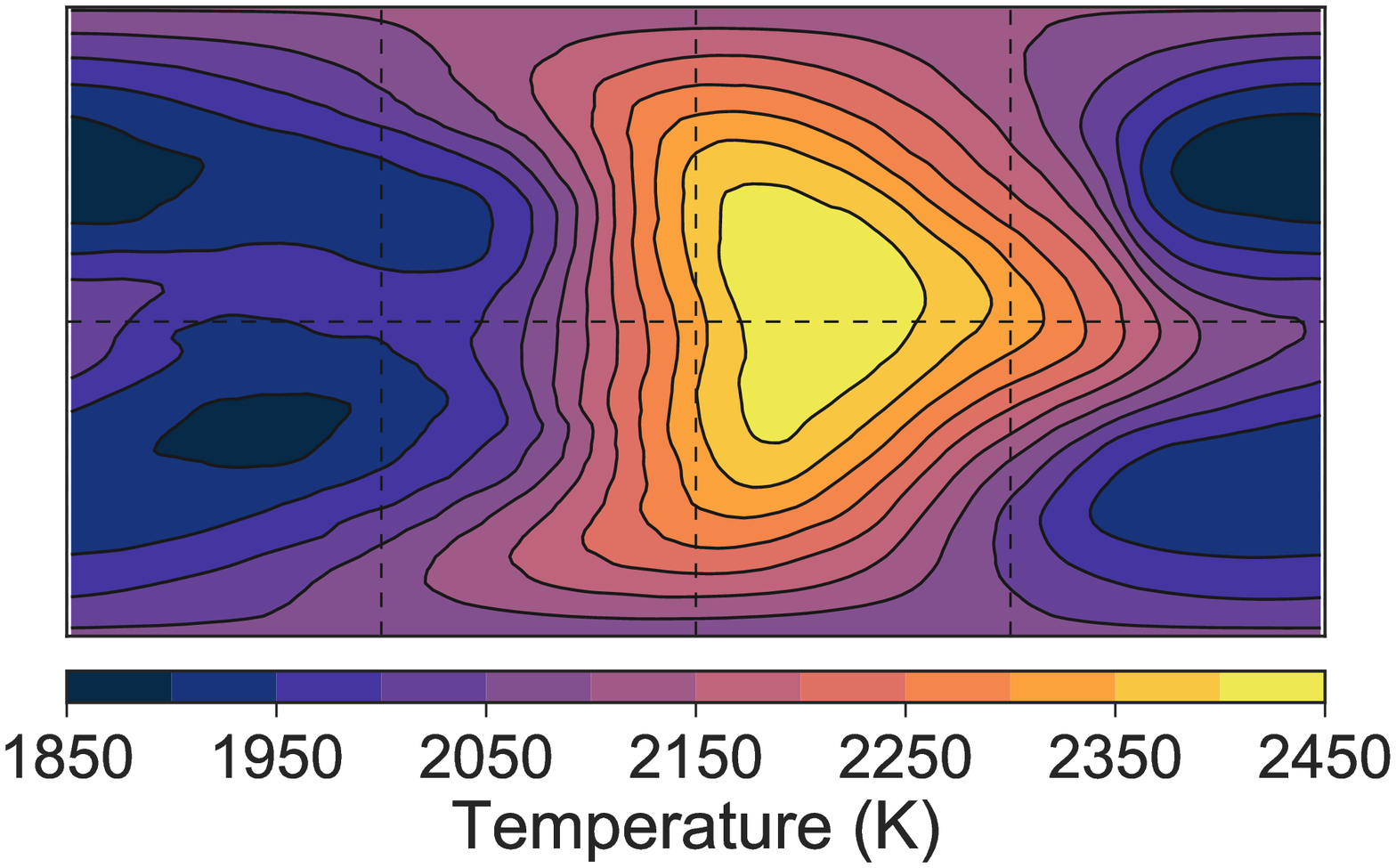}{0.3\textwidth}{$4.6\ \mathrm{gmol}^{-1}$, H\textsubscript{2}+N\textsubscript{2}, $\tau_{\infty} = 2.0$: half-surface-pressure air temperature}
           \fig{H2N2_5_halfp.eps}{0.3\textwidth}{$4.6\ \mathrm{gmol}^{-1}$, H\textsubscript{2}+N\textsubscript{2}, $\tau_{\infty} = 4.0$: half-surface-pressure air temperature}
           \fig{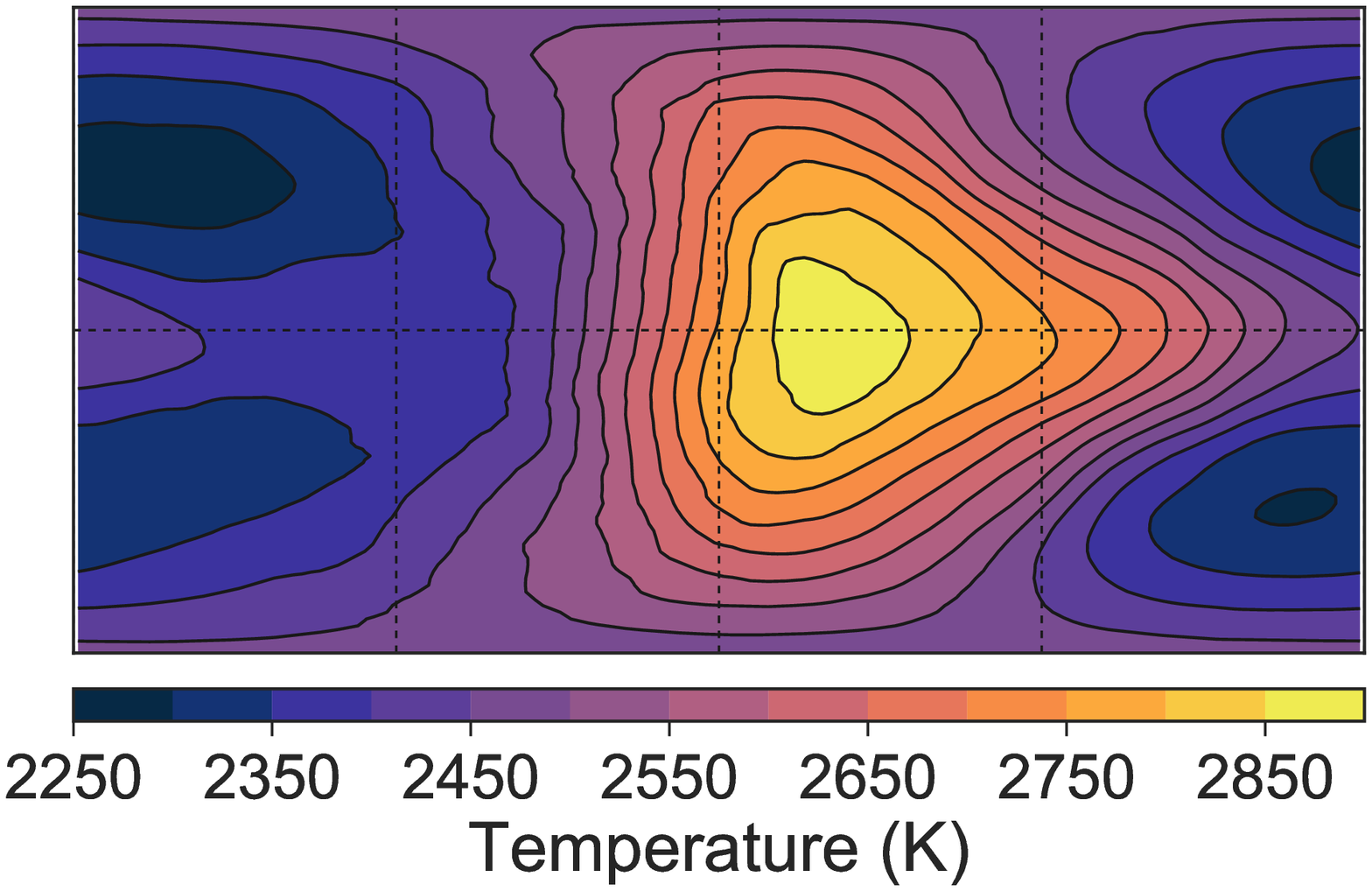}{0.3\textwidth}{$4.6\ \mathrm{gmol}^{-1}$, H\textsubscript{2}+N\textsubscript{2}, $\tau_{\infty} = 8.0$: half-surface-pressure air temperature}}

\caption{10-day average temperature maps of the half-surface-pressure level for $4.6\ \mathrm{gmol}^{-1}$  H\textsubscript{2} + N\textsubscript{2} atmospheres with surface pressure 5 bar and optical thicknesses of 2.0, 4.0, and 8.0.\label{fig:vary_tau_maps}}
\end{figure*}

The previous section tested the effect of changing the surface pressure. We kept the atmospheric composition and opacity the same, but this meant that the optical thickness $\tau_{\infty}$ changed with the surface pressure. In this section, we test the effect of changing the optical thickness with all other parameters constant, and demonstrate that the optical thickness does not greatly affect the global circulation, temperature distribution, and phase curve as discussed in Section \ref{sec:theory}.

We modelled three 5 bar $4.6\ \mathrm{gmol}^{-1}$ atmospheres with $\tau_{\infty} =$ 8.0, 4.0, and 2.0. Section \ref{sec:theory} predicts that the optical thickness will not have a large effect on the global circulation and temperature distribution (other than on the magnitude of the temperatures). Figure \ref{fig:vary_tau_maps} shows the temperature at the half surface pressure level for these three tests. All the tests have a similar global temperature distribution, and the tests with higher optical thickness have higher temperatures as expected. Figure \ref{fig:phasecurves_tauinf} in the next section shows that $\tau_{\infty}$ only affects the magnitude of the thermal phase curves of these tests.

The test with $\tau_{\infty} = 2.0$ is not hot enough to match the observations. The test with $\tau_{\infty} = 8.0$ matches the peak of the observed phase curve better than the $\tau_{\infty} = 4.0$ case, but its night-side is much hotter than the observations. If the criterion was only to match the magnitude and position of the peak of the phase curve, the $\tau_{\infty} = 8.0$ case would be our ``best-fit''. However, in Section \ref{sec:condensables} we show that the $\tau_{\infty} = 4.0$ case could also match the night-side observations given high night-side cloud formation, so we choose this to be our ``best-fit'' case.

%%%%%%%%%%%%%%%%%%%%%%%%%%%%%%%%%%%%%%%%%%%%%%%%%%%%%%%%%%%%%%%%%%%%%%%%%%

\subsection{Vertical Structure}\label{sec:vertical_structure}
\begin{figure*}
  \gridline{\fig{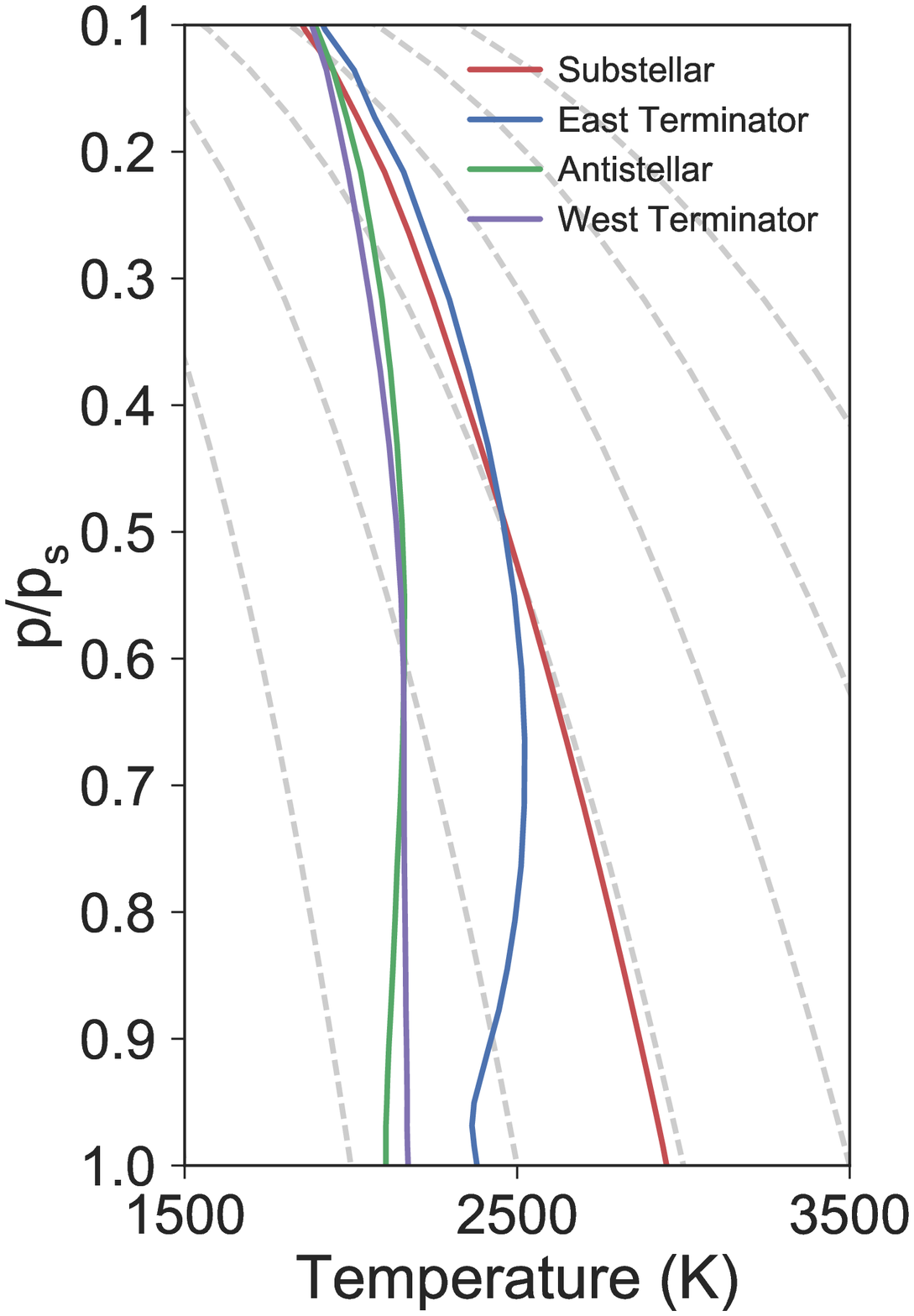}{0.3\textwidth}{(a) Temperature profile for several columns around the equator of the 5 bar $4.6\ \mathrm{gmol}^{-1}$ H\textsubscript{2}+N\textsubscript{2} atmosphere, with adiabats plotted in grey.}
            \fig{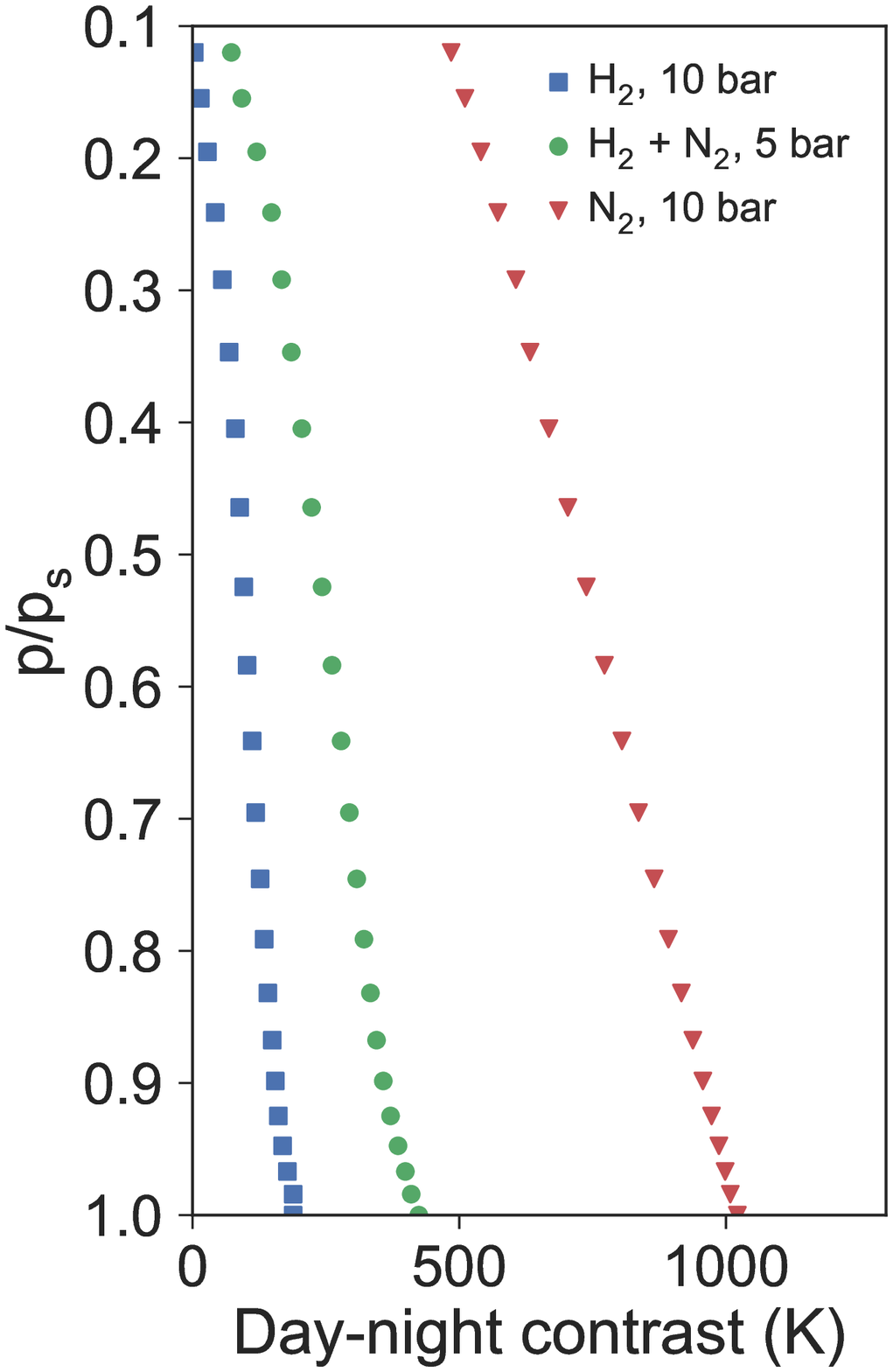}{0.3\textwidth}{(b) Day-night contrast versus pressure level, defined as the difference between the warmest and coolest hemispheres for consistency with \citet{demory2016map}.}
            \fig{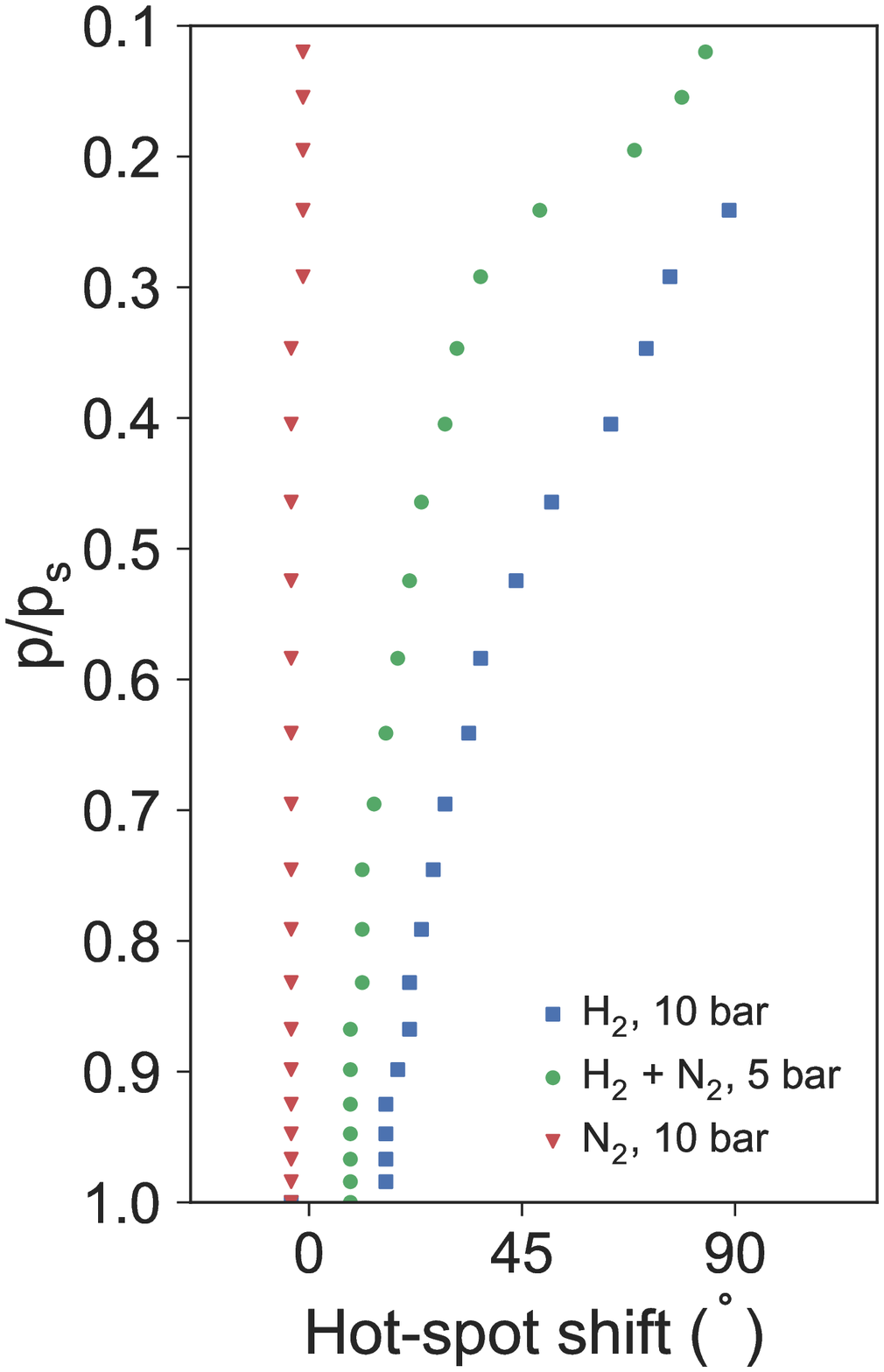}{0.3\textwidth}{(c) Hot-spot shift versus pressure level for all atmospheres, up to a level where the zonal temperature gradient is too small for a distinct hot-spot shift.}}
\caption{Vertical structure of the tested atmospheres. The best-fit ``H\textsubscript{2}+N\textsubscript{2}, 5 bar'' case here has mean molecular weight $4.6\ \mathrm{gmol}^{-1}$ and $\tau_{\infty} = 4.0$. The T(p) profiles follow the dry adiabat on the lower atmosphere of the day-side, but tend towards isothermal elsewhere. The day-night contrast always increases closer to the surface. The hot-spot shift generally increases further from the surface, but becomes indistinct without a large day-night contrast. \label{fig:pressure_variation}}
\end{figure*}

The brightness temperature measured in a thermal phase curve depends on the composition via two sets of properties. The atmospheric composition affects the radiative features which determine the heating rate, and the thermodynamic parameters affect the temperature distribution with latitude, longitude, and pressure. Finally the radiative features determine how this manifests as a brightness temperature to an observer. In this section we will discuss how the important features of the temperature distribution vary with depth in the atmosphere, and consider how the effect of the thermodynamic parameters can be distinguished from the effect of the radiative features on observations.

Figure \ref{fig:pressure_variation}a shows the temperature-pressure profiles of evenly spaced vertical columns around the equator of the planet. The planet is heated at the substellar point, where air rises through the deep convective troposphere. The global circulation discussed in Section \ref{sec:theory} moves heat eastwards in the mid-atmosphere, generating an inversion at the east terminator due to rapid surface cooling there. This shows the importance of a GCM over 1D models to this investigation, as some profiles are greatly perturbed from radiative-convective equilibrium by the atmospheric dynamics. It also shows how features such as the day-night contrast vary with depth, as the difference between the maximum and minimum temperatures is much greater at the surface.

Figure \ref{fig:pressure_variation}b shows that the day-night contrast is largest low in the atmosphere. This suggests that the large observed day-night brightness temperature contrast of 1300 K may be due to emission from the lower atmosphere of the planet. This could be explained a greenhouse gas with a generally high longwave opacity to account for the high temperatures, but with a window at $4.5\ \micron$ so the radiating level is low in the atmosphere at this wavelength.

Figure \ref{fig:pressure_variation}c shows how the hot-spot shift varies with pressure level in the atmosphere. It increases with height because the heat transport is stronger higher in the atmosphere. The lower atmosphere is also strongly coupled to the surface temperature, which tends towards no hot-spot shift due to the distribution of the incoming shortwave radiation.

\begin{figure}
\plotone{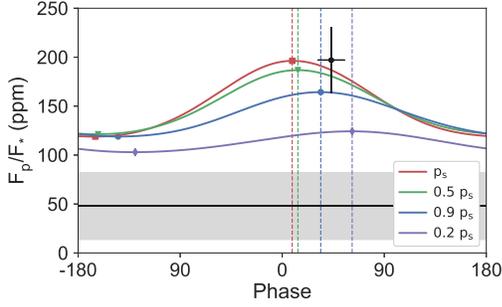}
\caption{Phase curves for different radiating levels in the 5 bar, $4.6\ \mathrm{gmol}^{-1}$,  H\textsubscript{2} + N\textsubscript{2} atmosphere. Raising the radiating level has a very similar effect to decreasing the molar mass in Figures \ref{fig:phasecurves_flux} and \ref{fig:phasecurves_temp}, leading to a degeneracy in interpreting the observed phase curve (the black point and line show the maximum and minimum observed fluxes, with error bars).\label{fig:phasecurves}}
\end{figure}

Figure \ref{fig:phasecurves} shows how the radiating level determines the observed phase curve due to the vertical structure of the atmosphere. The different phase curves correspond to the $4.5\ \micron$ emission from various pressure levels in the 5 bar, H\textsubscript{2} + N\textsubscript{2} atmosphere. The phase curve of the lower atmosphere has a much larger amplitude than the upper atmosphere, while the peak offset is much larger in the upper atmosphere. This corresponds to a larger day-night contrast in the lower atmosphere, and a larger hot-spot shift in the upper atmosphere. The grey-gas model OLR corresponds roughly to the 0.2 p\textsubscript{s} radiating level (see Section \ref{sec:sim_pc}).

The fundamental degeneracy in the observations is made clear by comparing Figure \ref{fig:phasecurves} and Figure \ref{fig:phasecurves_flux}. Both show a family of curves which vary from a large amplitude curve centered on 0$\degr$, to a small amplitude curve with a large offset. Increasing $t_{rad}/t_{transp}$ by changing the composition has the same effect as observing a radiating level higher in the atmosphere.

Observations at different wavelengths could resolve these degeneracies by  probing different pressure levels. For instance, it is possible that the high brightness temperatures observed are due to weaker atmospheric absorption at $4.5\ \micron$ than the average longwave opacity, so the measured radiation is from a lower level than the grey-gas OLR. H\textsubscript{2}-H\textsubscript{2} or H\textsubscript{2}-N\textsubscript{2} collision-induced absorption could cause this effect, as it has weaker absorption at $4.5\ \micron$ than on average in the thermal infrared at these temperatures \citep{wordsworth2013hydrogen}. H\textsubscript{2}O could also fill this role, as it has strong overall thermal infrared absorption but weak absorption at $4.5\ \micron$ (unless it is abundant enough for self-induced continuum absorption). CO\textsubscript{2} and CO do absorb in this region, so could not be abundant if our suggestion is correct. Broadband observations would measure the overall longwave radiating level, and could be compared to the $4.5\ \micron$ measurements to identify an absorption window.

%%%%%%%%%%%%%%%%%%%%%%%%%%%%%%%%%%%%%%%%%%%%%%%%%%%%%%%%%%%%%%%%%%%%%%%%%%

\subsection{Simulating Phase Curves}\label{sec:sim_pc}

We simulated the $4.5\ \micron$ phase curves of each test to directly compare them to the observations of 55 Cnc e, and to make their amplitude and peak offset clear. These are more practically useful than the temperature maps in section \ref{sec:results}, which show the results of the atmospheric dynamics but do not quantitatively show the amplitude and peak offset that would be observed.

The $4.5\ \micron$ phase curve was calculated using the weighted outgoing $4.5\ \micron$ spectral radiance (from the outgoing grey-gas flux brightness temperature). We integrated over the hemisphere centered on each grid cell around the equator in turn \citep{cowan2008inverting}:

\begin{equation}
  I_{p}(\xi) = \frac{\int_{-\pi/2}^{\pi/2} \int_{-\xi-\pi/2}^{-\xi+\pi/2}I_{4.5}^{\uparrow}|_{p=0}\cos(\lambda+\xi)\cos^{2}(\theta)d \lambda d \theta}{\int_{-\pi/2}^{\pi/2} \int_{-\xi-\pi/2}^{-\xi+\pi/2}\cos(\lambda+\xi)\cos^{2}(\theta)d \lambda d \theta}
\end{equation}

for phase angle $\xi$, outgoing $4.5\ \micron$ flux $I_{4.5}^{\uparrow}|_{p=0}$, longitude $\lambda$, and latitude $\theta$.

The planetary flux F\textsubscript{p} is compared to the stellar flux F\textsubscript{$\circledast$} \citep{crossfield2012acme}:

\begin{equation}
  \frac{F_{p}}{F_{\circledast}} = \frac{I_{p}}{I_{\circledast}}\Big( \frac{r_{p}}{r_{\circledast}} \Big) ^{2}
\end{equation}

for $\frac{r_{p}}{r_{\circledast}} = 0.0187$, and I\textsubscript{$\circledast$} given an effective temperature of 5196 K \citep{von201155}.

\begin{figure}
\plotone{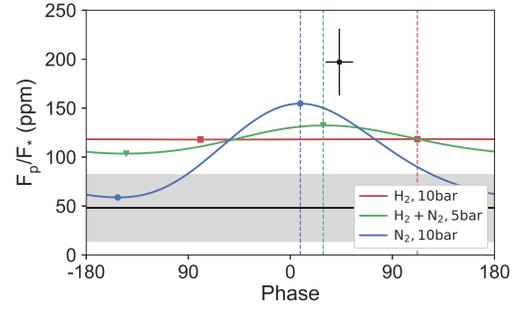}
\caption{Simulated $4.5\ \micron$ phase curves calculated from the brightness temperature of the grey-gas OLR. The red curve is the 10 bar H\textsubscript{2} atmosphere, which has such efficient heat transport that it has a large peak offset and very small amplitude. The blue curve is the 10 bar N\textsubscript{2} atmosphere, with very weak heat transport so a large amplitude and peak offset. The green curve is the 5 bar H\textsubscript{2}+N\textsubscript{2} atmosphere, with a significant offset and amplitude. The offset and amplitude are not as large as the \citet{demory2016map} measurements, shown by the black point and line (with their errors shown by the bars and the shaded area).\label{fig:phasecurves_flux}}
\end{figure}

Figure \ref{fig:phasecurves_flux} shows the resulting phase curves for different mean molecular weight values. The maximum and minimum fluxes of the real, observed phase curve are plotted as points. It is clear that the single-gas tests do not fit the data well. Test 1 (H\textsubscript{2}, 10 bar, $\tau_{inf} = 8.0$) has a hot-spot shift, but a very flat curve due to its efficient heat circulation. Test 2 (H\textsubscript{2}, 10 bar, $\tau_{inf} = 8.0$) has a large amplitude, but no hot-spot shift due to its small radiative timescale. The phase curve of Test 5 (H\textsubscript{2}+N\textsubscript{2}, 5 bar, $\tau_{inf} = 4.0$) fits the observations better, as does its temperature distribution in section \ref{sec:theory}. It has a large peak offset and amplitude, although not as large as the observed offset and amplitude. This discrepancy may be partly due to the high radiating level imposed by the grey-gas approximation with a high optical thickness.

\begin{figure}
\plotone{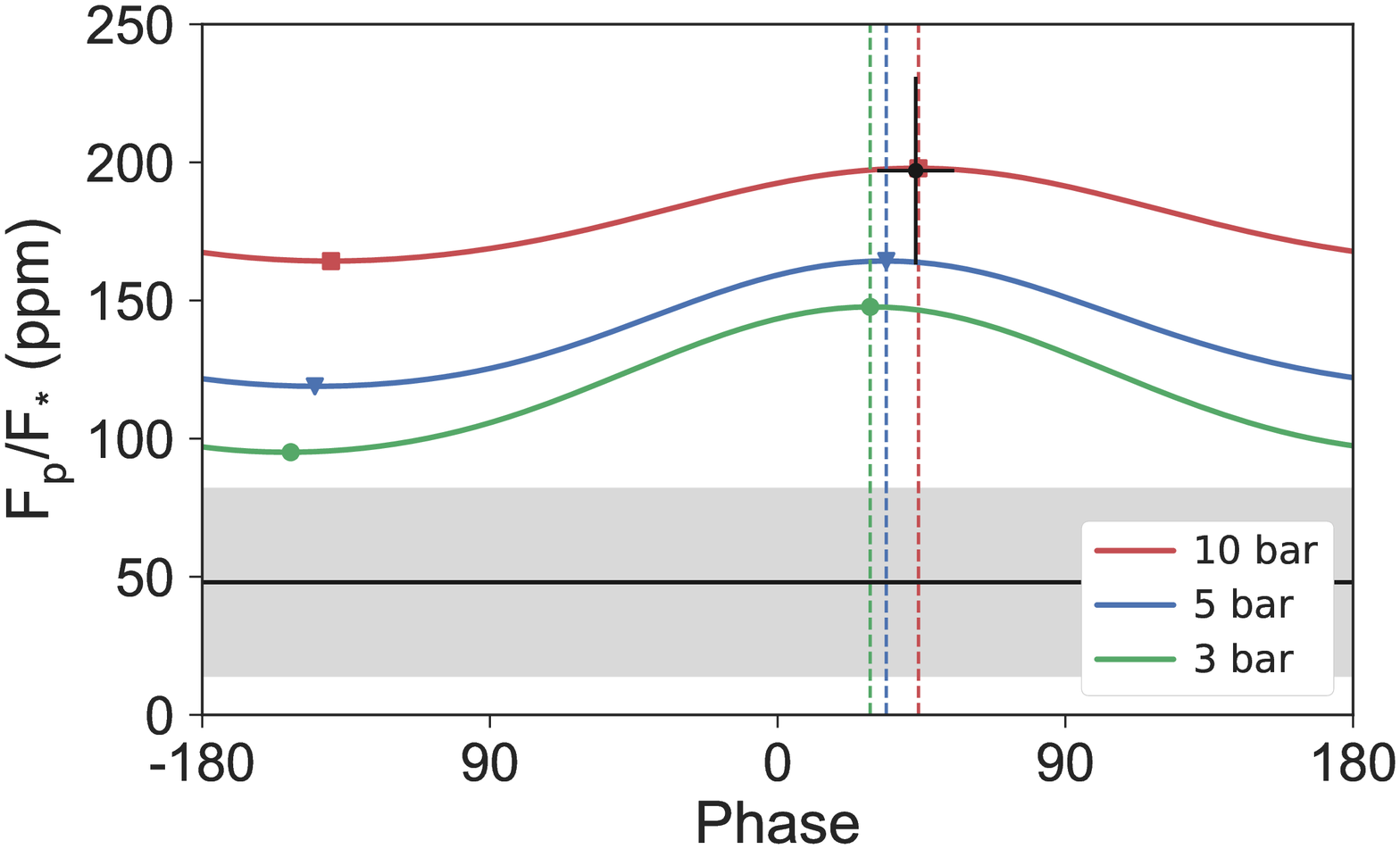}
\caption{Phase curves calculated using the emission from the half-surface-pressure level of the $4.6\ \mathrm{gmol}^{-1}$ H\textsubscript{2} + N\textsubscript{2} atmospheres with surface pressures of 3, 5, and 10 bar, corresponding to the temperature maps in Figure \ref{fig:H2N2_T_maps}.\label{fig:phasecurves_H2N2}}
\end{figure}

\begin{figure}
\plotone{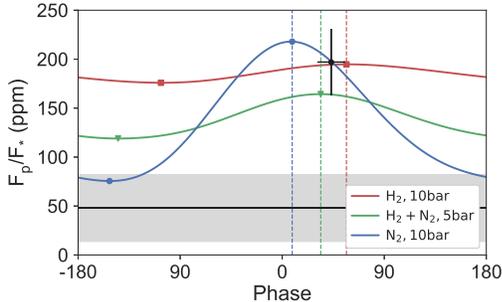}
\caption{Simulated phase curves for the emission from a radiating level at half-surface-pressure. The amplitude and offset are larger than the phase curves of the OLR. The offset and amplitude are not as large as the \citet{demory2016map} measurements, but Figure \ref{fig:phasecurves_clouds} shows that the H\textsubscript{2}+N\textsubscript{2} atmosphere (blue curve) could match the observations with the night-side cloud formation discussed in Section \ref{sec:condensables}.\label{fig:phasecurves_temp}}
\end{figure}

Figure \ref{fig:phasecurves_temp} shows the phase curves which would be measured from a radiating level at half the surface pressure for each mean molecular weight test -- a possibility if the atmosphere's opacity at $4.5\ \micron$ is lower than its mean opacity. The H\textsubscript{2}-N\textsubscript{2} test fits the observations better in this figure, with a larger phase curve peak offset and amplitude than the OLR phase curve. The night-side flux is still too high, but this could be explained by cloud formation on the night-side (see Section \ref{sec:condensables}).

We varied the surface pressure of the 4.6 gmol\textsuperscript{-1} H\textsubscript{2}-N\textsubscript{2} case, to determine the ``best-fit'' to the observed phase curve (discussed in Section \ref{sec:ps_effect}). Figure \ref{fig:phasecurves_H2N2} shows the phase curves of the emission from the half-pressure level for the three tests. These show how increasing the pressure increases the offset and peak magnitude, but decreases the amplitude (as the opacity was constant, so the optical thickness increased). If the criterion were just to match the position and magnitude of the hot-spot shift, the 10 bar case would be the best fit. However, we chose the 5 bar case as our ``best-fit'' as it has the possibility to match the observations given high cloud formation on the night-side, which we discuss in Section \ref{sec:condensables}.

\begin{figure}
\plotone{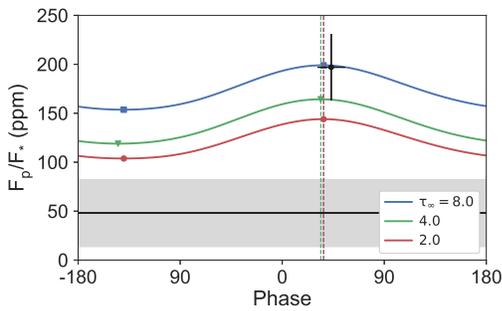}
\caption{Phase curves calculated using the emission from the half-surface-pressure level of the 5 bar $4.6\ \mathrm{gmol}^{-1}$ H\textsubscript{2} + N\textsubscript{2} atmospheres with optical thicknesses of 2.0, 4.0, and 8.0, corresponding to the temperature maps in Figure \ref{fig:vary_tau_maps}.\label{fig:phasecurves_tauinf}}
\end{figure}

Figure \ref{fig:phasecurves_tauinf} shows the phase curves of the tests in Section \ref{sec:tauinf_effect}, where we varied the optical thickness of the 5 bar 4.6 gmol\textsuperscript{-1} H\textsubscript{2}-N\textsubscript{2} to determine its effect on the global circulation and temperature distribution. As expected, the optical thickness does not affect the global temperature distribution and phase curve, beyond the magnitude of the temperatures and fluxes.

In general, the phase curves calculated using the temperature of the half-pressure level matched observations better than those from the grey-gas OLR. The hypothetical absorption window at $4.5\ \micron$ discussed above could be responsible, and would also explain the high temperatures observed on the day-side. These require a high mean longwave opacity to maintain high surface temperature but a low $4.5\ \micron$ opacity to allow radiation from the hot layers to escape to space and account for the observed 4.5 $\micron$ brightness temperature.

To summarise, the thermal emission from the half-pressure level of the 5 bar, 4.6 gmol\textsuperscript{-1} H\textsubscript{2}+N\textsubscript{2} test matched the observed phase curve peak offset and magnitude in Figure \ref{fig:phasecurves_temp}, but did not match the minimum flux (night-side temperature). None of the phase curves calculated with the model grey-gas OLR matched the observations, but we do not consider this to be important -- in reality, the radiating level of the $4.5\ \micron$ emission will depend entirely on the radiative species in the atmosphere. We discuss the effect of clouds and condensables in more detail in the next section, as they could decrease the apparent night-side temperature and better explain the observations.

%%%%%%%%%%%%%%%%%%%%%%%%%%%%%%%%%%%%%%%%%%%%%%%%%%%%%%%%%%%%%%%%%%%%%%%%%%

\subsection{Condensables and Clouds}\label{sec:condensables}

It is possible that clouds form high on the colder night-side so the photosphere there is higher and cooler. This would lead to less thermal emission from the night-side, and a larger day-night contrast and phase curve amplitude. \citep{parmentier2016transitions}. In this section, we use a simple calculation to estimate the effect of night-side clouds on the phase curve of our 4.6 gmol\textsuperscript{-1}, 5 bar H\textsubscript{2}+N\textsubscript{2} atmosphere, which was the test with  the coolest night-side which also matched the observed phase curve peak amplitude and offset.

Clouds could be formed by condensables such as SiO or Na from a day-side magma ocean. \citet{miguel2011compositions} calculated the partial pressures of different species over magma in a vacuum at different temperatures, and showed that a magma ocean at a temperature around 2700 K (the measured mean day-side temperature) would support a significant partial pressure of multiple species, the most abundant being SiO and Na which would both have partial pressures of approximately 10 mbar. The maximum surface temperatures of our tests are over 3000 K, where SiO becomes more abundant and reaches partial pressures of hundreds of mbar.

We decided to focus on clouds at the top of the atmosphere to find the upper limit on the possible effect of clouds, as in \citet{parmentier2016transitions}. We used \citet{miguel2011compositions} to estimate a range of surface partial pressures of SiO and Na, based on the surface temperature of the hot-spot. Then, we calculated the saturation partial pressure of SiO and Na for the highest level of each column \citep{wetzel2013sio}. If this was larger than the surface partial pressure, we set the radiating level to the top of the atmosphere in that column, and recalculated the phase curve.

\begin{figure}
\plotone{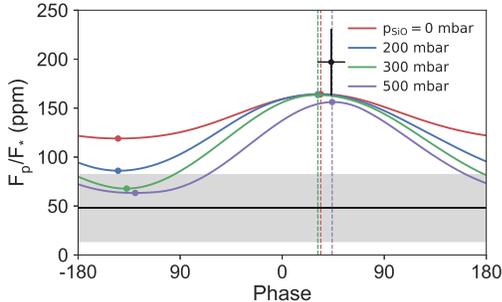}
\caption{Simulated phase curves for different equilibrium surface partial pressures of SiO in the 4.6 gmol\textsuperscript{-1}, 5 bar H\textsubscript{2} + N\textsubscript{2} atmosphere. 10 mbar corresponds to the mean day-side surface temperature, 50 mbar to the mean temperature in the region between 25N to 25S and 25E to 25W, and 100 mbar to the maximum surface temperature. The 300 mbar curve shows that clouds could form on the day-side at high enough surface partial pressures.  The offset and amplitude of the 100 mbar case almost agrees with the \citet{demory2016map} measurements within error. It is not clear which partial pressure is appropriate, as the calculated pressures strictly apply to magma under a vacuum \citep{miguel2011compositions}, and there is not a single clear magma ocean temperature in our results. \label{fig:phasecurves_clouds}}
\end{figure}

In the range of partial pressures from \citet{miguel2011compositions}, we found that SiO could condense on the night-side of some of our tests, but that Na would not condense in any tests. Figure \ref{fig:phasecurves_clouds} shows that at high enough equilibrium partial pressures, the SiO clouds could significantly increase the day-night contrast and phase curve amplitude. For a partial pressure of 300 mbar, the new phase curve matches the observations of \citet{demory2016map} within error. Figure \ref{fig:phasecurves_clouds} also shows that at high SiO partial pressures, heterogeneous day-side cloud formation can increase the hot-spot shift, as clouds tend to form towards the cooler western terminator on the day-side \citep{parmentier2016transitions}. This effect is small in our modelled atmospheres but might be important for different atmospheric compositions or different condensables.

Further observations at more wavelengths would be needed to find clouds on the planet, or to discover which species are present. Measuring and understanding the condensables present could help to break the degeneracies discussed in Section \ref{sec:vertical_structure}, as their concentrations could be linked to the real surface temperature via the calculations of \citet{miguel2011compositions}. These calculations also only apply to partial pressures in a vacuum, so further work on this topic would benefit from the partial pressures outgassed into an atmosphere from a magma ocean.

In conclusion, SiO from a magma ocean is a good candidate for a cloud species which would affect observations. It could form clouds high on the night-side which would increase the phase curve amplitude, making our 4.6 gmol\textsuperscript{-1}, 5 bar H\textsubscript{2} + N\textsubscript{2} test consistent with the observations. Future work could include a model of cloud formation and transport, which would be more realistic than our post-processing which essentially only predicts cloud formation in cold areas of the atmosphere.

\section{Discussion}\label{sec:discussion}
We can use our theory and results from sections \ref{sec:theory} and \ref{sec:results} to make suggestions about the composition of an atmosphere on 55 Cnc e that would produce the observed thermal phase curve. It is important to reiterate our main assumptions that the atmosphere is composed primarily of diatomic molecules and has negligible shortwave opacity. Further modelling work should investigate the effect of real-gas radiation and variable molar heat capacity.

Our best fit to the observations was the 90\%-10\% mixture of H\textsubscript{2} and N\textsubscript{2} with a mean-molecular weight of 4.6 gmol\textsuperscript{-1}, a specific heat capacity of 7443 Jkg\textsuperscript{-1}K\textsuperscript{-1}, optical thickness 4.0, and a 5 bar surface pressure. Figure \ref{fig:param_map} uses the theory of \citet{zhang2016effects} to show that any atmosphere with a significant hot-spot shift and day-night contrast must have a similar composition to this.

The phase curve shown in Figure \ref{fig:phasecurves_flux} from the OLR of this best-fitting test did not match the measured phase curve of \citet{demory2016map}, as its offset and amplitude were too small. The other tests could reproduce one of these features to be large enough, but not both at the same time. The phase curve calculated from the thermal emission of the half-pressure level in Figure \ref{fig:phasecurves_temp} matched the observed phase curve peak offset and amplitude, but did not match the minimum (without night-side cloud formation).

This difference could be due to a number of things. Firstly, the model and theory might be inaccurate or overly simplified. Secondly, our parameter space of simple H\textsubscript{2} and N\textsubscript{2} atmospheres may not represent the real atmosphere. It is still possible that the main component could be a gas such as CO\textsubscript{2} with a different molar heat capacity, which is the main variable we have not investigated. Very high or low surface pressures are also possible, and might require a different modelling approach (for example, a volatile-dominated thin atmosphere).

However, we can explain the difference via other physical processes. We have discussed the effect of clouds formed by condensables, which could form on the night-side and greatly increase the phase-curve amplitude. These condensables could also affect the vertical structure and dynamics. There may be behaviour which is not captured by our grey-gas approximation, where optically active species might absorb in the shortwave or let through $4.5\ \micron$ flux, which would affect the atmospheric structure and dynamics, and its phase curve.

Tidal heating could help to explain the observed day-side and night-side temperatures. 55 Cnc e is expected to have an eccentricity of approximately 0.001, leading to tidal heating between $10^{-3}\ \mathrm{Wm}^{-2}$ and $10^{6}\ \mathrm{Wm}^{-2}$ (\citet{bolmont2013tidal}, \citet{demory2016map}). Without an atmosphere, a flux of $10^{6}\ \mathrm{Wm}^{-2}$ would raise the substellar temperature by 200 K, partly explaining the day-side temperature.

The 1300 K night-side temperature could also be sustained by a tidal heating flux of approximately $10^{5}\ \mathrm{Wm}^{-2}$, removing the need for an atmosphere to keep the night-side warm. However, in order for such a large flux to diffuse through a solid rock layer with typical diffusivity would require the crust to be a mere 3mm thick. A solid crust would not be stable under such circumstances, implying a global magma ocean with consequently high night-side temperature. In addition, tidal heating does not explain the observed hot-spot shift. We can speculate, however, that tidal heating could play a role in explaining the magnitude of the observed day-side temperature, especially if transport of tidal heating to the surface favoured the partially molten day-side over the cooler night-side.

Therefore, we suggest that an atmosphere is the current best explanation for the observations. We can use our results to rule out certain atmospheres and make suggestions about a likely atmospheric composition. A single-gas, clear-sky atmosphere with only H\textsubscript{2} or N\textsubscript{2} is not consistent with the observations, as their thermodynamic properties preclude both a large phase curve amplitude and peak shift.

Our suggested climate is therefore our ``best-fit'' atmosphere with: mean-molecular weight of 4.6 gmol\textsuperscript{-1}, specific heat capacity of 7443 Jkg\textsuperscript{-1}K\textsuperscript{-1}, optical thickness of 4.0, and 5 bar surface pressure. This could fit the observed phase curve given high night-side clouds and a window at $4.5\ \micron$. The composition could be a 90\%-10\% mixture of H\textsubscript{2} and N\textsubscript{2} with some trace greenhouse gases such as CO\textsubscript{2} or H\textsubscript{2}O, and with cloud-forming species such as SiO from a day-side magma ocean.

%%%%%%%%%%%%%%%%%%%%%%%%%%%%%%%%%%%%%%%%%%%%%%%%%%%%%%%%%%%%%%%%%%%%%%%%%%

\section{Conclusions}\label{sec:conclusions}

The large hot-spot shift and day-night temperature contrast of the thermal phase curve of 55 Cnc e present a puzzle. We used theories of circulation on tidally locked planets to predict the composition of simple atmospheres on 55 Cnc e which would show either of these features. We modelled these climates, which qualitatively agreed with our predictions. We then predicted and modelled a ``best-fit'' composition, which had a significant hot-spot shift and day-side contrast. This atmosphere did not match the measured phase curve by itself, but could match it given high night-side clouds and an absorption window at $4.5\ \micron$. We showed that the global circulation, temperature distribution, and thermal phase curve depends strongly on the mean molecular weight and the surface pressure. The atmospheric optical thickness and opacity does not greatly affect the global temperature distribution beyond its magnitude.

Further modelling work should include the effects of real-gas radiation, condensables, and clouds. These could all affect the atmosphere's horizontal and vertical structure, as well as the radiating level and outgoing radiation. Shortwave absorption from atmospheric gases and clouds, heat transport by condensables, and the effect of scattering could be important to the atmospheric structure and circulation, and to the observed phase curve.

Observations at different wavelengths would be invaluable in breaking the degeneracies described in this paper. Broadband observations could reveal the overall brightness temperature of the planet, and answer the questions we have raised about a spectral window at $4.5\ \micron$. Observations at other wavelengths could probe different levels of the atmosphere and be compared to the vertical structure of models such as that shown in Section \ref{sec:results}. These would help to solve the degeneracies between radiating level and composition that we discussed above. Further, the indications of an $\mathrm{H_2}$-rich atmosphere from our fit to the phase curve are problematic in view of the likely high H\textsubscript{2} escape rate from 55 Cnc e.  It should be noted that it is only the hot spot phase shift that pushes the fit toward a low molecular weight atmosphere; other features could be accounted for with a high molecular weight atmosphere dominated by, e.g., $\mathrm{N_2}$, $\mathrm{CO}$, or $\mathrm{CO_2}$. For future observations of this planet and other lava planets, accurate determination of the hot-spot shift is essential.

In this paper we explored the range of surface pressures from 3 to 10 bar. Very thin atmospheres (including thin rock-vapour atmospheres such as posited by \citep{castan2011hot}) are inconsistent with the fairly large observed night-side temperature.  However, 55 Cnc e could conceivably have a much thicker atmosphere than we investigated. For an atmosphere with surface pressure of hundreds or thousands of bars, the infrared opacity of any plausible constituent would make the brightness temperature insensitive to circulation and temperature deep in the atmosphere, except insofar as the deep circulation affects the circulation in the upper atmosphere. If the upper atmosphere were to decouple dynamically from the deep atmosphere, the phase curve could become independent of surface pressure.  A very massive atmosphere, however, would tend to have very high surface temperatures, and the effects on vapourisation from a probably global magma ocean might have observable consequences.  The observational signature of massive atmospheres on lava planets constitutes a fruitful area for future study. This regime would require attention to the effects of atmospheric absorption of incoming stellar energy, neglected in the present study.

Some of our test runs showed transience on the scale of days in their temperature distributions, which might be detectable by future observations with high time resolution. These seemed to be caused by the cold-spot cyclones moving around the planets, and the varying jet speed and position. The 5 bar H\textsubscript{2}+N\textsubscript{2} atmosphere varied its $4.5\ \micron$ flux phase curve amplitude by 20\% and its offset by 30\% (10$\degr$) over the course of ten days. This is significantly less than the variation of 400\% reported by \citet{demory2016variability}, but this short-period variability could be a target for future observations. For comparison, the phase curve amplitude of the H\textsubscript{2} 10 bar test only varied by a few percent over the same time period. Observations of similar variability have already been made of a hot giant planet by \citet{armstrong2016variability}.

We set out to test whether the observed phase curve is inconsistent with the presence of an atmosphere more massive than the thin rock-vapour atmospheres typically assumed for lava planets. We have shown that an atmosphere with 5 bar surface pressure and a mean molecular weight of $4.6\ \mathrm{gmol}^{-1}$ could be consistent with the observations given cloud formation high on the night-side. We hope that this paper has demonstrated how the climates of terrestrial planets can be constrained by models working closely with observations, and how suggestions can be made about the composition of their atmospheres. Similar studies should become possible for many more planets as observational and modelling capabilities improve.

\acknowledgments

M.H. was supported by an STFC studentship. We thank the anonymous reviewer for their time and valuable comments, which greatly improved the paper.

\end{document}